\documentclass[prb,twocolumn,notitlepage,eqsecnum,nofootinbib,floatfix]{revtex4-1}
\usepackage{graphicx,amsmath,amssymb,bm,comment}
\usepackage{enumerate}
\usepackage{mathtools}
\usepackage{subcaption}

\makeatletter
\newcommand*{\citenumns}[2][]{%
  \begingroup
  \let\NAT@mbox=\mbox
  \let\@cite\NAT@citenum
  \let\NAT@space\NAT@spacechar
  \let\NAT@super@kern\relax
  \renewcommand\NAT@open{}%
  \renewcommand\NAT@close{}%
  \cite[#1]{#2}%
  \endgroup
}
\makeatother

\begin{document}

\title{Six textbook mistakes in data analysis}
\author{Alexandros Gezerlis}
\affiliation{Department of Physics, University of Guelph, Guelph, Ontario N1G 2W1, Canada}
\author{Martin Williams}
\affiliation{Office of Teaching and Learning \& Department of Physics, University of Guelph, Guelph, Ontario N1G 2W1, Canada}

\date{\today}

\begin{abstract}
This article discusses a number of incorrect statements appearing in textbooks on data analysis, machine learning, or computational methods; the common theme in 
all these cases is the relevance and application of statistics to the study of scientific or engineering data; these mistakes are also quite prevalent in the research literature. 
Crucially, we do not address errors made by an individual author, focusing instead on mistakes that are widespread in the introductory literature. 
After some background on frequentist and Bayesian linear regression, we turn to our 
six paradigmatic cases, providing in each instance a specific example of the textbook mistake, pointers to the specialist literature where the topic is handled properly,
along with a correction
that summarizes the salient points.
The mistakes (and corrections) 
are broadly relevant
to any technical setting where statistical techniques are used
to draw practical conclusions, ranging from topics introduced
in an elementary course on experimental measurements all the way to more involved approaches
to regression.
\end{abstract}

\maketitle

\section{Introduction}

In an earlier work we lamented the presence of many mistaken
claims in computational-physics textbooks.~\cite{GezerlisWilliams}
As noted there, while such errors are
understandable in the research literature (where things are still rapidly evolving), 
they are more problematic (and pernicious) in the context of introductory textbooks. The present article
is of a similar nature, in that it addresses a number of widespread misconceptions;
this time we focus on textbook coverage of \textit{data analysis}, a term typically used to describe the way that  data is collected, processed, summarized, interpreted, etc. 
(Most of our discussion applies to linear statistical inference, but the lessons 
learned are of wider import.) These are topics that are taught
at the undergraduate 
level either in a dedicated 
course on experimental measurements, or as part of a course on numerical methods/computation.

Whether in traditional experimental analyses, or in more recent theoretical approaches,
or even using tools borrowed from machine learning/data science, 
the theme of uncertainty quantification is increasingly important in science \& engineering.
The topics discussed in the present article involve applications of statistics of varying sophistication;
given the broader community
and applicability this entails, it should not come as a surprise that many of
the mistakes discussed below are drawn from textbooks which have garnered tens of thousands of citations (each). We have selected these mistakes based on 
their being both widespread and important; each of the six incorrect claims elaborated
on below appears in at least two standard textbooks. In other words,
our focus is not on the (inevitable) typos or misunderstandings that arise when 
an individual author garbles a given topic, but on widespread errors that (while
known to experts) continue to propagate through the introductory literature. 

One of the authors recently finished updating an introductory 
computational-physics textbook,\cite{Gezerlis} a process preceded by a deep dive into the data-analysis/machine-learning literature.
The six themes discussed below are handled correctly in Ref.~\citenumns{Gezerlis}, 
but we
also cite many works written by statisticians, in the spirit of providing 
the reader with many reliable resources. Our intended audience consists mainly
of science or engineering instructors who are interested in data analysis; our hope is that by bringing
attention to these incorrect claims we can make sure that fewer students are exposed
to them in their undergraduate education in the future; similarly, the points 
we raise
may be of service to non-expert or novice researchers who currently treat some of these techniques as black boxes. 
Since many of these mistakes arise due to 
the use of inscrutable (or rarely scrutinized) mathematical symbols, it may be beneficial 
to start by discussing the notation.

\begin{figure*}[t]
\centering
   \begin{subfigure}{0.49\textwidth} \centering
     \includegraphics[scale=0.45]{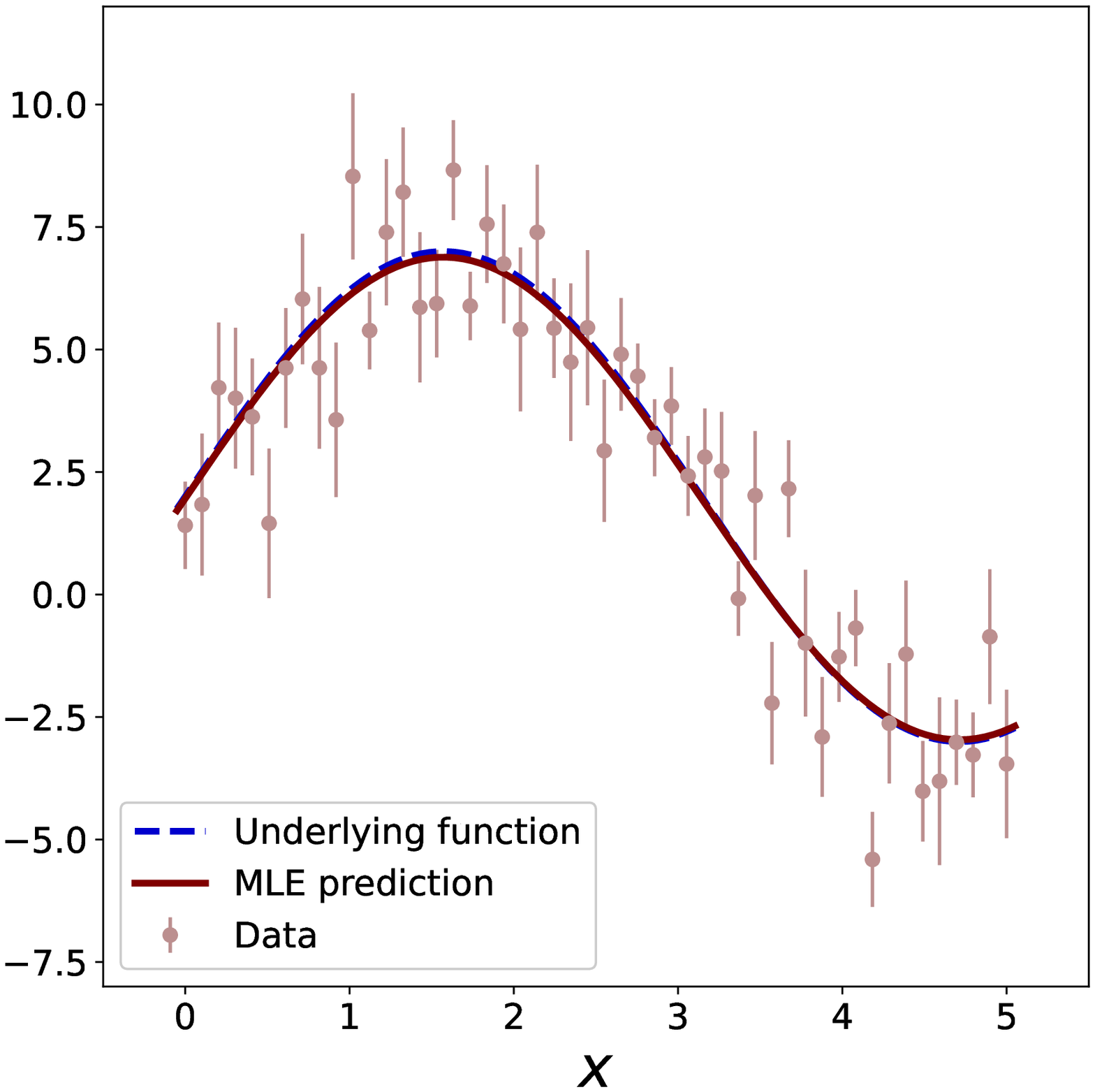}
     \caption{}
   \end{subfigure}
   \begin{subfigure}{0.49\textwidth} \centering
     \includegraphics[scale=0.45]{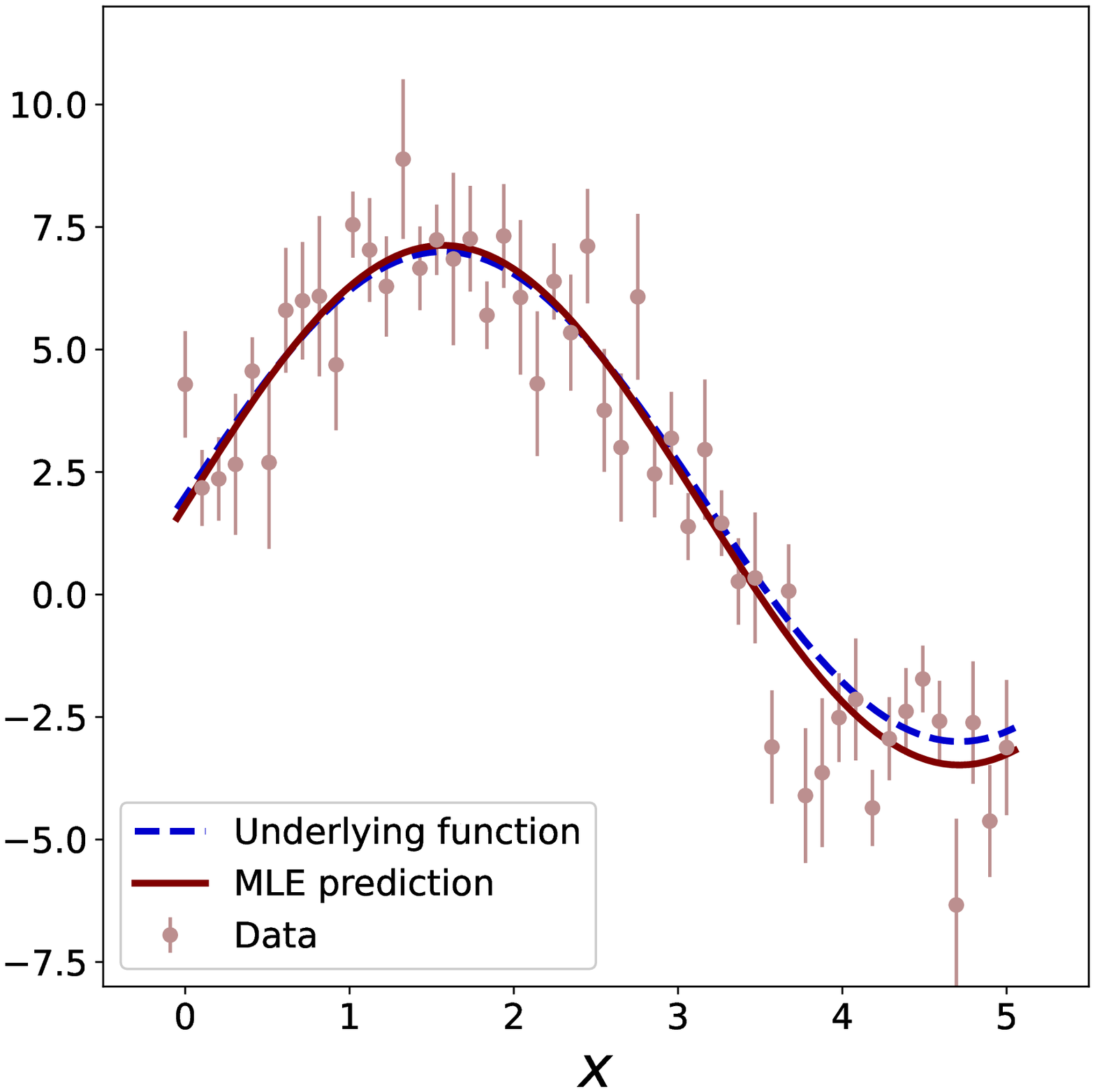}
     \caption{}
   \end{subfigure}
\caption{Starting from the underlying function $2 + 5 \sin x$, we 
add in different noise terms ${\cal E}_j$ (left and right) to produce synthetic data.
Each panel also shows the prediction corresponding to the 
maximum-likelihood point estimate.} \label{fig:rawdata}
\end{figure*}

\section{Establishing the notation}
\label{sec:notation}

Nearly all textbooks on numerical methods discuss the (linear algebra) problem 
of least-squares fitting: this is typically taken to involve as input a table
of values, $(x_j, y_j)$ where $j = 0, 1, \ldots, N-1$ and $N$ is the number of 
data points. In science it is more common to encounter, instead, the case
of \textit{heteroskedastic errors}, i.e., the scenario where each input data 
value $y_j$ is itself the result of a separate experiment (and therefore 
associated with an input-data uncertainty, $\sigma_j$). In other words,
our input data can be thought of as $(x_j, y_j \pm \sigma_j)$. Whatever 
form our approximating function may take, we know that it is expected
to somehow capture the behavior of the data $y_j$, while also 
taking into consideration the uncertainty $\sigma_j$ associated with each input data point.
From a linear-algebra perspective, this gives rise to \textit{weighted least squares};
from a statistical perspective, the same results are more commonly arrived at in the context
of \textit{maximum-likelihood estimation} (MLE)---to which we return below.

Since the notation here can, indeed, get messy/confusing, let us take the time
to carefully distinguish between the true underlying behavior and our dataset. (Our discussion is intended merely
to set the stage for section~\ref{sec:mistakes}, i.e., this is \textit{not} a full
pedagogical exposition.) 
We explicitly model the discrepancy between theory and experiment as follows:
\begin{equation}
Y_j = \sum_{k=0}^{n-1} c_{\star, k} \phi_{k}(x_j) + {\cal E}_j
\label{eq:Bayes1}
\end{equation}
Here $Y_j$ are the data (``dependent variable'') values, the $\phi_k$'s are (input/frozen) basis functions,
evaluated at the input (``independent variable'') values $x_j$, the $c_{\star, k}$'s are the true/underlying
parameter values, and ${\cal E}_j$ is a continuous random variable. Crucially,
here we are assuming that we know both the true theory (the $\phi_k$'s)
and the true parameter values (the $c_{\star, k}$'s); we will discuss below
what to do when (as is usual in practice) you don't actually know these. The way
to interpret Eq.~(\ref{eq:Bayes1}) is as adding \textit{noise} (the second term on the right-hand side) to the true/actual/theory
value (the first term on the right-hand side), thereby producing the experimental
value (on the left-hand side).

If we bundle together all the parameter values into a column vector and
all the basis functions into a row vector (i.e., give $k$ the values $0, 1, \ldots n-1$---where, crucially, $n \neq N$), Eq.~(\ref{eq:Bayes1}) can be compactly
stated in the form $Y_j = \bm{\phi}_j \mathbf{c}_{\star}  + {\cal E}_j$. (Similarly,
you could bundle together all $j$ values to summarize things as
$\mathbf{Y} = \mathbf{\Phi} \mathbf{c}_{\star} + \bm{{\cal E}}$.)
It's now time to make a specific assumption about the type of ${\cal E}_j$
we are faced with: we take this to be the distribution ${\cal N}(0, \sigma_j)$.
Here we are employing the (standard) notation for a normal/Gaussian probability density function:
\begin{equation}
{\cal N}(\mu, \sigma) = \frac{1}{\sqrt{2\pi} \sigma} \exp 
\left [ - \frac{1}{2} \left ( \frac{x - \mu}{\sigma} \right )^2 \right ]
\label{eq:Bayes3}
\end{equation}
where, as usual, the mean $\mu$ determines the location of the peak and the 
standard deviation $\sigma$ its width. (Note that, in this notation, ${\cal N}(\mu, \sigma)$ lists the standard deviation, not the variance.) 
As a result of the central limit theorem, asymptotic normality does appear quite often
in practice; even so, here we are explicitly taking the normality of ${\cal E}_j$ 
to be an assumption. 
Since ${\cal E}_j$
obeys the distribution ${\cal N}(0, \sigma_j)$, the data point $Y_j$ of Eq.~(\ref{eq:Bayes1}) obeys
the distribution ${\cal N}\left (\bm{\phi}_j \mathbf{c}_{\star}, \sigma_j \right )$.
Since each data point $Y_j$ obeys a different distribution,
we are dealing with independent but not identically
distributed samples. In Fig.~\ref{fig:rawdata} we show a couple of $N=50$ synthetic
data sets, generated by adding noise to the $n=2$ true theory with 
$\phi_0 = 1$, $\phi_1 = \sin x$ and $c_{\star, 0} = 2$, $c_{\star, 1} = 5$;
the two panels result from different noise terms, ${\cal E}_j$.

Up to this point, we have been addressing only 
the \textit{data-generating distribution}, corresponding to the true parameters 
$\mathbf{c}_{\star}$; since these are unknown in practice, we 
now turn to a discussion of how one might go about \textit{estimating} them, given
a dataset $\mathbf{y}$. (Note a notational choice here: the random vector
bundling together the possible input-data values is $\mathbf{Y}$, while
a concrete realization, i.e., a given dataset, is denoted by $\mathbf{y}$.) 
Two standard formulations of this problem are (a) maximum-likelihood estimation,
and (b) a Bayesian approach. Here we do not enter into the relative merits of
frequentist vs Bayesian approaches to statistics; most introductory textbooks
with a scientific or engineering audience are still based on a frequentist approach;
as a result, four out of the six mistakes discussed below arise in the context of
(implicitly) frequentist discussions. 
With that in mind, 
we now briefly summarize both a frequentist and a Bayesian approach to linear
regression.

Starting with the frequentist approach: since the data-generating distribution
giving rise to $Y_j$ in Eq.~(\ref{eq:Bayes1}) was taken to be normal/Gaussian, 
it is reasonable to assume that something similar holds for a general
parameter set $\mathbf{c}$:
\begin{equation}
P(\mathbf{y} ; \mathbf{c}, \bm{\Phi}, \bm{\Sigma}_d) 
=  \prod_{j=0}^{N-1} \frac{1}{\sqrt{2\pi} \sigma_j}  \exp 
 [ - \frac{1}{2} \sum_{j=0}^{N-1} \left ( \frac{y_j - \bm{\phi}_j \mathbf{c}}{\sigma_j} \right )^2  ]
\label{eq:Bayes6}
\end{equation}
where, crucially, this involves $\mathbf{c}$ instead of $\mathbf{c}_{\star}$;
here $\bm{\Sigma}_d$ bundles together the input-data variances (i.e., the $\sigma_j^2$'s).
Qualitatively, Eq.~(\ref{eq:Bayes6}) is a statistical model telling us how a 
given $\mathbf{c}$ gives rise to $\mathbf{Y}$, of which a given realization
(a dataset) is $\mathbf{y}$; holding everything else fixed, we can then view 
Eq.~(\ref{eq:Bayes6}) as a function of the parameters only: 
$L(\mathbf{c}) = P(\mathbf{y} ; \mathbf{c}, \bm{\Phi}, \bm{\Sigma}_d)$. This is 
known as the \textit{likelihood} of the parameters. Maximizing the likelihood
gives rise to the \textit{maximum-likelihood estimator} (MLE) $\hat{\mathbf{c}}$:
\begin{align}
\hat{\mathbf{c}} = (\bm{\Phi}^T \bm{\Sigma}_d^{-1} \bm{\Phi} )^{-1} \bm{\Phi}^T \bm{\Sigma}_d^{-1}\mathbf{Y}, \qquad \hat{\bm{\Sigma}}_{kk} 
= \left ( \bm{\Phi}^T \bm{\Sigma}_d^{-1} \bm{\Phi} \right )^{-1}_{kk}
\label{eq:ls18_proba}
\end{align}
where we took the opportunity to also show the variances in the MLE parameter
estimates, $\hat{\bm{\Sigma}}_{kk}$. (As above, we make the distinction
between the MLE estimator involving $\mathbf{Y}$ and the MLE point estimate 
involving $\mathbf{y}$.) In Fig.~\ref{fig:rawdata} we also show
the prediction corresponding to the maximum-likelihood parameter estimate 
$\hat{\mathbf{c}}$ (for each dataset), i.e.,  
$\sum_{k=0}^{n-1} \hat{c}_{k} \phi_{k}(x)$; observe that Eq.~(\ref{eq:Bayes6})
and Eq.~(\ref{eq:ls18_proba}) know nothing of the underlying function (involving
the true parameters 
$\mathbf{c}_{\star}$), i.e.,
they take as input only a given dataset (each time). Take a moment to appreciate the
fact that how closely
the MLE prediction matches the underlying function will also
depend on the noise term(s) that each panel corresponds to.
Qualitatively, the motivation behind maximum-likelihood 
estimation is to pick that set of parameters which makes the experimentally
determined dataset most probable; as we further discuss below, two major complications
arise in this connection, first, $N$ might be small and, second, we typically
have access to only a single dataset $\mathbf{y}$.

The Bayesian outlook combines the statistical model of how the parameter values
give rise to the dataset (i.e., the likelihood) with all prior information 
one has about the parameters themselves; crucially, now the parameters are
a random vector, $\mathbf{C}$. Combining the likelihood and the prior is accomplished via the application
of \textit{Bayes' rule}:
\begin{align}
P(\mathbf{c} | \mathbf{y}; \bm{\Phi}, \bm{\Sigma}_d, \bm{\mu}_0, \bm{\Sigma}_0) &= 
\frac{P(\mathbf{y} | \mathbf{c}; \bm{\Phi}, \bm{\Sigma}_d) P(\mathbf{c} ; \bm{\mu}_0, \bm{\Sigma}_0)}{P(\mathbf{y} ; \bm{\Phi}, \bm{\Sigma}_d, \bm{\mu}_0, \bm{\Sigma}_0)} \nonumber \\
&= \frac{P(\mathbf{y} | \mathbf{c}; \bm{\Phi}, \bm{\Sigma}_d) P(\mathbf{c} ; \bm{\mu}_0, \bm{\Sigma}_0)}{\int d^n c
P(\mathbf{y} | \mathbf{c}; \bm{\Phi}, \bm{\Sigma}_d) P(\mathbf{c} ; \bm{\mu}_0, \bm{\Sigma}_0)}
\label{eq:Bayesrule3}
\end{align}
The likelihood is now written as $P(\mathbf{y} | \mathbf{c}; \bm{\Phi}, \bm{\Sigma}_d)$; note that we are limiting ourselves to the single dataset $\mathbf{y}$.
The \textit{prior distribution} $P(\mathbf{c} ; \bm{\mu}_0, \bm{\Sigma}_0)$ is here
kept general, characterized by a mean vector $\bm{\mu}_0$ and a covariance matrix
$\bm{\Sigma}_0$. The denominator $P(\mathbf{y} ; \bm{\Phi}, \bm{\Sigma}_d, \bm{\mu}_0, \bm{\Sigma}_0)$ is the \textit{marginal likelihood}
(also known as the \textit{evidence}) and plays an important role in model selection;
here we take it to be simply a normalization factor.
Note that $P(\mathbf{y} ; \bm{\Phi}, \bm{\Sigma}_d, \bm{\mu}_0, \bm{\Sigma}_0)$ does
\textit{not} depend on the parameter values $\mathbf{c}$: as clearly shown
in the second equality, these are integrated out. The main entity in
Bayesian regression is the \textit{posterior distribution}  
$P(\mathbf{c} | \mathbf{y}; \bm{\Phi}, \bm{\Sigma}_d, \bm{\mu}_0, \bm{\Sigma}_0)$,
shown on the left-hand side of Eq.~(\ref{eq:Bayesrule3}). It encapsulates
all that we know about the parameter values, combining both prior knowledge
and what we learned from the dataset itself. 
Without getting into a detailed analysis of possible prior and likelihood forms, 
we note a standard textbook example, that where all three relevant entities
(prior, likelihood, posterior) are Gaussian functions, leading to the posterior
covariance matrix and mean vector:
\begin{equation}
\bm{\Sigma}_{\mathbf{c}} = \left ( \bm{\Phi}^T \bm{\Sigma}_d^{-1} \bm{\Phi} 
+ \bm{\Sigma}_0 ^{-1} \right )^{-1}, 
\bm{\mu}_{\mathbf{c}} = 
 \bm{\Sigma}_{\mathbf{c}}
\left ( \bm{\Phi}^T \bm{\Sigma}_d^{-1} \mathbf{y} 
 + \bm{\Sigma}_0^{-1} \bm{\mu}_0 \right )
 \label{eq:Baynew6}
\end{equation}
This can be viewed as a generalization of Eq.~(\ref{eq:ls18_proba}) such that
the prior's $\bm{\mu}_0$ and $\bm{\Sigma}_0$ are taken into account.
One attractive feature of the Bayesian approach to linear regression is that,
having computed the posterior distribution $P(\mathbf{c} | \mathbf{y}; \bm{\Phi}, \bm{\Sigma}_d, \bm{\mu}_0, \bm{\Sigma}_0)$, one may then proceed to ``fold it in''
to evaluate other quantities. An important example involves carrying out \textit{predictions}, i.e., determining the $\tilde{y}$ value corresponding
to a new $\tilde{x}$ value; crucially, these $\tilde{x}$ and $\tilde{y}$ 
are \textit{not} part of the dataset. In the Bayesian spirit, the right approach
is to write down a \textit{posterior predictive distribution}:
\begin{align}
&P(\tilde{y} | \mathbf{y}; \tilde{\bm{\phi}}, \bm{\Phi}, \bm{\Sigma}_d, \bm{\mu}_0, \bm{\Sigma}_0) = \mathbb{E} [ P(\tilde{y} | \mathbf{c}; \tilde{\bm{\phi}}) ] \nonumber \\
&= \int d^n c   P(\tilde{y} | \mathbf{c}; \tilde{\bm{\phi}})  P(\mathbf{c} | \mathbf{y}; \bm{\Phi}, \bm{\Sigma}_d, \bm{\mu}_0, \bm{\Sigma}_0)
\label{eq:predi1}
\end{align}
which is here seen to be an expectation of our predictive model
with respect to the posterior (as spelled out in the second equality);
here $\tilde{\bm{\phi}}$ bundles together the $\phi_k$'s evaluated at $\tilde{x}$, 
i.e., $\phi_k(\tilde{x})$ for all $k$'s.

\section{Mistakes and corrections}
\label{sec:mistakes}

Our aim in the present work is to focus on the correct understanding of 
data analysis, not to criticize the authors of respected textbooks; with that
in mind, we now cite a superset of references, made up of many standard textbooks covering the relevant material.\cite{Barlow,Beers,Bertsekas,Bevington,Bishop,Bohm,Boudreau, Burden, Chapra, Degroot, Deisenroth, 
DeVries, Gilat, Gould, Hamming, Jiang, Kahaner, Kiusalaas, Koonin, Landau, Lyons, Mandel, Mathews, Murphy, Press, Pruneau, Rice, Roe, Rogers, Sirca, Sivia, Theodoridis, Thompson, Wong,Zielesny} 
(We have provided the Editor and Referees of the present manuscript
with detailed bibliographic information.) 
For each of the six themes we discuss below, our organizing principle will
be to first provide some conceptual and contextual background, then give the \textit{mistake} (namely, a specific quote from the literature),
followed by a discussion of why the quote is substantively wrong (typically
using equations and/or a figure), and conclude with a \textit{correction} (namely,
an improved few-sentence formulation, intended to supersede the mistake).
Since this is a short article, it is inevitable that some of the more involved
misconceptions are touched upon but not deeply explored; 
to remedy this, we provide several pointers to the specialist literature in what follows.
In most of the quotes given below we have tweaked the mathematical notation
(following section~\ref{sec:notation})
in order to make the discussion cohere across all topics; we have not modified
the original quotes in any other way (unless so marked).

\subsection{Maximum-likelihood parameter estimation}
\label{sec:MLE}

In the process of introducing a new technique, an instructor/textbook author is expected
to provide some motivation; a good outcome is that a newly introduced 
approach feels intuitive to students soon after they hear about it/see it applied. 
In this connection, a danger that every instructor must guard against is going
so far as to make the new approach seem self-evident: if the technique is self-evident,
then there is (a) no need to consider (the possibility of) alternative approaches,
(b) a risk that students who fail to see why the technique is logically obvious end
up blaming themselves. These considerations are quite relevant to the way 
maximum-likelihood estimation is often introduced in textbooks:

{\bf Mistake \#1} ``\textit{We assume that the observed set of measurements is more
likely to have come from the parent distribution [corresponding to $\mathbf{c}_{\star}$]
than from any other similar distribution with different coefficients and, therefore,
the probability [sic] 
\begin{equation}
P(\mathbf{y} ; \mathbf{c}_{\star}, \bm{\Phi}, \bm{\Sigma}_d) 
= \prod_{j=0}^{N-1} \frac{1}{\sqrt{2\pi} \sigma_j}  \exp 
 [ - \frac{1}{2} \sum_{j=0}^{N-1} \left ( \frac{y_j - \bm{\phi}_j \mathbf{c}_{\star}}{\sigma_j} \right )^2 ]
\label{eq:parent}
\end{equation}
is the maximum probability [sic] attainable with Eq.~(\ref{eq:Bayes6}). Thus, the maximum-likelihood estimates for $\mathbf{c}$ are those values that maximize the probability [sic]
of Eq.~(\ref{eq:Bayes6}).}''

As implied above, in such expositions
the attempt to justify likelihood maximization has gone too far: to make Mistake \#1
is to look at Eq.~(\ref{eq:parent}), notice its formal similarity to Eq.~(\ref{eq:Bayes6}), and thereby draw the conclusion that 
the value of $\mathbf{c}$ that maximizes the probability density in Eq.~(\ref{eq:Bayes6}) gives the true parameters $\mathbf{c}_{\star}$.
(Incidentally, Eq.~(\ref{eq:Bayes6}) and therefore also Eq.~(\ref{eq:parent}),
are giving probability \textit{densities}, since $\mathbf{Y}$ is a continuous random vector.)
If maximum-likelihood estimation 
is so self-evident, that immediately raises the question why
other approaches to parametric inference (e.g., the method of moments) also exist:
if finding the argument that maximizes the likelihood gives you the true parameters,
then why would you not simply go ahead and just find the true parameters? 
A point that will keep recurring in our discussion below: when such strong assumptions
are smuggled into textbooks (especially those of a ``practical'' flavor)
students are left either scratching their heads or pretending that they get it.

Let us elaborate a bit on why this misconception arose: if you look at 
Eq.~(\ref{eq:Bayes6})---keeping the $y_j$'s, the $\sigma_j$'s, and the $\bm{\phi}_j$'s fixed---then a bit of 
thought will convince you that, indeed, the only way to get the answer of 
Eq.~(\ref{eq:parent}) is precisely to set $\mathbf{c} = \mathbf{c}_{\star}$ in 
Eq.~(\ref{eq:Bayes6}). Note, however, that this fact is unrelated to the 
\textit{maximization} of the likelihood: indeed, if we knew the true parameters
we could plug them into the likelihood 
$L(\mathbf{c}) = P(\mathbf{y} ; \mathbf{c}, \bm{\Phi}, \bm{\Sigma}_d)$
and thereby get the parent distribution, $P(\mathbf{y} ; \mathbf{c}_{\star}, \bm{\Phi}, \bm{\Sigma}_d)$ but this line of thinking still doesn't tell us how to go about
approximating $\mathbf{c}_{\star}$ in the first place; just because you have 
a Gaussian, doesn't mean it's centered at the right place. The confusion seems
to arise from a number of interrelated facts. First, a Gaussian distribution has a single, well-defined maximum, so it is enticing to think that the (special) location of that 
maximum should somehow be related to quantity we are after. Second, it is 
qualitatively plausible to want to maximize the likelihood: since the random 
vector $\mathbf{Y}$ did give rise to the specific dataset $\mathbf{y}$, shouldn't we
want to make it probable that the dataset that did arise would arise?
Third, likelihood maximization as per Eq.~(\ref{eq:ls18_proba}) leads to several
pleasing properties, e.g., in the asymptotic limit (i.e., as $N \rightarrow \infty$)
the MLE estimate is consistent (i.e., $\hat{\mathbf{c}}$ goes to $\mathbf{c}_{\star}$),
the MLE estimator is unbiased (i.e., $\mathbb{E}(\hat{\mathbf{c}}) = \mathbf{c}_{\star}$, 
where this frequentist expectation is across datasets),
and $\hat{\mathbf{c}}$ minimizes the (Kullback--Leibler)
 distance between the general likelihood $P(\mathbf{y} ; \mathbf{c}, \bm{\Phi}, \bm{\Sigma}_d)$ and the parent distribution $P(\mathbf{y} ; \mathbf{c}_{\star}, \bm{\Phi}, \bm{\Sigma}_d)$; see Refs.~\citenumns{Casella}~and~\citenumns{Wasserman} for more details (and more properties).

\begin{figure}[t]
\centering
\includegraphics[trim={0cm 2.5cm 0 3.5cm},clip,width=0.99\columnwidth]{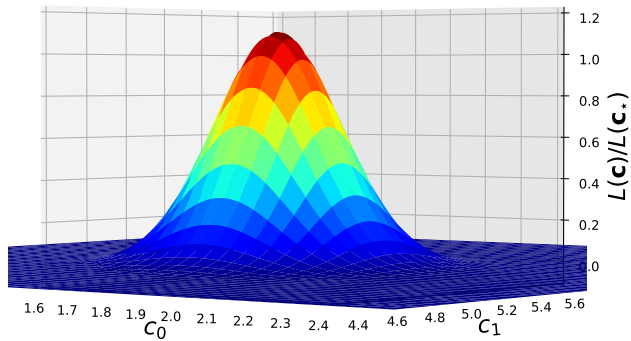}
\caption{Likelihood function $L(\mathbf{c})$ for a given
dataset $\mathbf{y}$ and a two-dimensional parameter set $\mathbf{c}$,
normalized by the likelihood's value at the 
true parameters $\mathbf{c}_{\star}$.} 
\label{fig:like}
\end{figure}

The nice properties of MLE discussed in the previous paragraph referred either
to the asymptotic limit or to integrations across all possible datasets. Let us, instead,
examine a single dataset with a finite number of points, $N$; for concreteness,
take the $N=50$ dataset from the left panel of Fig.~\ref{fig:rawdata}. 
For this example,
we generated the dataset from a true/underlying function ourselves, so we can 
easily check whether or not the location of the maximum likelihood gives us the
true parameters or not.
The corresponding likelihood is plotted in Fig.~\ref{fig:like}, where you can clearly see
that at the maximum $L(\hat{\mathbf{c}}) > L(\mathbf{c}_{\star})$ holds,\cite{Kendall} in direct
contradiction of the quote we saw above, which assumed that $L(\hat{\mathbf{c}}) = L(\mathbf{c}_{\star})$. (For the sake of completeness, we note that there exist
(non-Gaussian) likelihoods for which the 
maximum is not unique, or doesn't even exist; nobody would have taken it for granted
in those cases that the $\mathbf{c}_{\star}$ are easy to find.) As a matter of fact,
the MLE method itself is telling us how much (or whether) to trust its
estimate $\hat{\mathbf{c}}$: we discuss this more below but, for now,
note that the point estimate is associated with a measure of its uncertainty
in Eq.~(\ref{eq:ls18_proba}).
To summarize the main point:

{\bf Correction \#1} \textit{Maximum-likelihood estimation is a reasonable
approach to parametric inference (as a matter of fact, it's the most common one).
While MLE obeys several nice properties, you should
not take it for granted that it automatically leads to the true/underlying parameter
values. This will be true only if your dataset size $N$ is huge or 
you are interested in average properties (across datasets). For a single dataset
of non-huge size $N$, you should carefully investigate the uncertainty in your
estimate. 
}

\subsection{Chi-squared statistic and quality of fit}
\label{sec:chisq}

The approach taken in section~\ref{sec:MLE}, MLE, is the statistical
foundation of the more prosaic technique introduced in nearly all computational-science
textbooks, namely least-squares fitting. The former models the discrepancy between
theory and experiment, trying to come up with systematic point estimates and 
uncertainties; the latter is usually provided in an \textit{ad hoc} format, 
i.e., not really justified. When the weighted least-squares approach is justified
in such works,
this is typically done hand-wavingly, precisely by bringing up maximum likelihood. 
The fact that the statistical background either goes unmentioned or is laconically
referred to can be detrimental to students' understanding of what is a 
random variable, what isn't, and why this matters. To make things concrete, 
let us turn to the next misconception (probably the most popular of the lot):

{\bf Mistake \#2} ``\textit{With this result in mind a statistical quantity named $\chi^2_{\text{red}}$ (`reduced chi-square') can 
be defined as
\begin{equation}
\chi^2_{\text{red}} = \frac{\chi^2}{N-n} = \frac{1}{N-n}\sum_{j=0}^{N-1} \left ( \frac{y_j - \bm{\phi}_j \mathbf{c}}{\sigma_j} \right )^2
\label{eq:ls3_again2}
\end{equation}
which evaluates to a value close to $1$ for a good fit since the number of data $N$ should be considerably larger than the number of parameters of the model function $n$, i.e. $N \gg n$.}''

Let us start by interpreting the $\chi^2$ statistic appearing in 
Eq.~(\ref{eq:ls3_again2}). In numerical-methods textbooks this is typically
motivated, reasonably enough, by noting that it attempts to minimize the distance 
between the experimental value, $y_j$, and the model prediction, $\bm{\phi}_j \mathbf{c}$,
weighted by the input-data uncertainty each time, $\sigma_j$. 
While there are other choices one could have made instead of least-squares
minimization (e.g., minimizing the maximum error), the above is certainly a 
plausible choice. Turning now to an actual justification of the form of
Eq.~(\ref{eq:ls3_again2}): taking the natural logarithm of 
$L(\mathbf{c}) = P(\mathbf{y} ; \mathbf{c}, \bm{\Phi}, \bm{\Sigma}_d)$
from Eq.~(\ref{eq:Bayes6}), we realize that all the prefactors are constant 
(since they are experimentally determined); since the exponential is monotonic,
to maximize $L(\mathbf{c})$ we can simply minimize what's in the exponent. But
the exponent is (within a factor of 2) precisely the $\chi^2$ statistic of 
Eq.~(\ref{eq:ls3_again2})! In other words, $\chi^2$ minimization is equivalent
to maximum-likelihood parameter estimation when the errors are normally distributed.
(Of course, MLE is much more general, since it also applies to cases 
where the errors are \textit{not} normally distributed.) Similarly,
the MLE solution of Eq.~(\ref{eq:ls18_proba}) is equivalent to the solution
of the normal equations appearing in the weighted least-squares problem.

So far, so good; the $\chi^2$ statistic appearing in Eq.~(\ref{eq:ls3_again2})
is either qualitatively motivated or explicitly derived from MLE but, either way,
its form is the same. This, however, is not the point of providing the above quote:
Mistake \#2 claims that a good fit is synonymous with $\chi^2 \approx N- n$.
First, some terminology: the difference between the number of data points $N$
and the number of parameters $n$ (when $\bm{\Sigma}_d$ is diagonal
and the $\phi_k$'s are linearly independent) 
is called the number of degrees of freedom, $\nu = N - n$. 
In other words, the frequently propagated advice is that $\chi^2_{\text{red}} = \chi^2/\nu$ should
be as close as possible (from above) to $1$; some expositions
also bring up (without justification) the incomplete gamma function at this point.
It is easy to see why such an approach
is so popular: for a given dataset and a given theory/model, Eq.~(\ref{eq:ls3_again2})
allows one (even one who is unfamiliar with modern statistics) 
to produce a single number and thereby draw the conclusion whether or
not the fit is good. As the authors of Ref.~\citenumns{Vehtari} put it:
``The most pernicious idea in statistics is the idea that we can produce a single-number summary of
any data set and this will be enough to make a decision.''


Before we turn to a general discussion of
what's wrong with this specific single-number summary,
let's look at a couple of examples. For 
the dataset from the left panel of Fig.~\ref{fig:rawdata},
the maximum-likelihood estimate leads to a (minimum) $\chi^2/\nu$ 
of $1.28$, whereas the dataset from the right panel 
corresponds to an MLE parameter set with a $\chi^2/\nu$ 
of $1.23$. This is already looking a little suspicious: 
the left panel exhibits a closer match between MLE prediction and the underlying
function, yet the value of $\chi^2/\nu$ is larger. Of course, in practical applications
the dataset (and the MLE prediction and corresponding $\chi^2/\nu$ value) are all
that we have, since the true functional form and the values of the true parameters
are unknown. Since both datasets were generated from the same true theory, one would be
ill-advised to try to determine which of the two models (of the same functional form)
is correct.

\begin{figure}[t]
\centering
\includegraphics[width=0.49\textwidth]{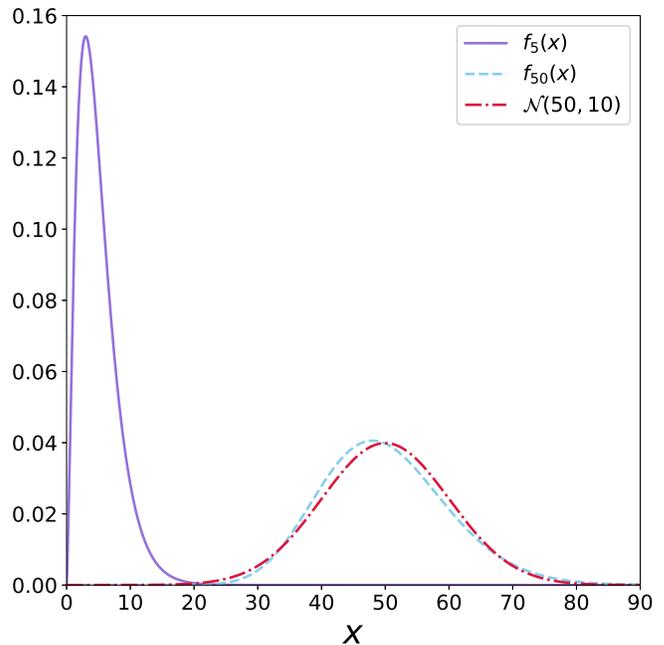}
\caption{Chi-squared distribution $f_{\nu}(x)$ for 5 and 50 degrees of freedom;
in the latter case, the corresponding asymptotic normal distribution is also shown.} 
\label{fig:chi}
\end{figure}

Let us now discuss the origin of the advice that for a good fit the
$\chi^2/\nu$ should
be as close as possible (from above) to $1$. 
First, we recall that the exponent in Eq.~(\ref{eq:Bayes6}) 
involves $Y_j$ and is therefore a continuous random variable. 
Setting $\mathbf{c} = \mathbf{c}_{\star}$, we realize that the numerator
is nothing other than the error term 
${\cal E}_j$,
as you can see from Eq.~(\ref{eq:Bayes1}). To see what this means,
examine the variance of the whole fraction (before squaring):
\begin{equation}
\mathbb{V} \left ( \frac{{\cal E}_j}{\sigma_j} \right ) = 
\frac{1}{\sigma_j^2} \mathbb{V}({\cal E}_j) = \frac{\sigma_j^2}{\sigma_j^2} = 1
\label{eq:Bayes7}
\end{equation}
We first used a 
standard property of the variance
and then realized that ${\cal E}_j$ obeys the distribution ${\cal N}(0, \sigma_j)$, so its variance is simply $\sigma_j^2$; we then cancelled, 
reaching the conclusion that each
term appearing in the exponent of Eq.~(\ref{eq:Bayes6}) and therefore also in the sum for $\chi^2$ 
has unit variance. Since 
${\cal E}_j$ had zero mean, so will ${\cal E}_j/\sigma_j$: in other words, we have shown
that ${\cal E}_j/\sigma_j$ obeys the distribution ${\cal N}(0, 1)$, i.e.,
it is a \textit{standard normal} random variable. An important lemma 
(Ref.~\citenumns{Casella}, p. 219) states that 
a random variable\index{random variable} that is produced by squaring $\nu$ independent 
standard normal random variables (and then adding the squares together) obeys
the \textit{chi-squared distribution}:
\begin{equation}
f_{\nu}(x) = \frac{1}{2^{\nu/2} \Gamma(\nu/2)} e^{-x/2} x^{(\nu/2) - 1}
\label{eq:chisqdistrib0}
\end{equation}
where $\Gamma$ is the \textit{gamma function} (unsurprisingly).
In Fig.~\ref{fig:chi} we are plotting this distribution
for a couple of values of $\nu$: for small $\nu$ the curve is characterized 
by considerable
skewness, but then becomes increasingly Gaussian-like as $\nu$ is increased. 

Crucially, the
$\chi^2$ of Eq.~(\ref{eq:ls3_again2}) obeys the 
distribution of Eq.~(\ref{eq:chisqdistrib0}); 
this explains why the statistic\index{statistic} and the distribution
have the same name. To emphasize the point: $\chi^2$ is a random variable,
characterized by a distribution
with a mean, variance, etc. Specifically, the distribution with the probability-density
function of Eq.~(\ref{eq:chisqdistrib0}) has a mean of $\nu$: 
this is precisely the source of the standard advice that a good fit means
$\chi^2/\nu \approx 1$. Here's the catch: this distribution has a variance 
of $2\nu$, so it is most certainly possible that a good fit will not 
give rise to a $\chi^2/\nu$ value that is equal to the mean of this distribution. 
Thus, one is not limited to the oft-repeated advice on how to interpret
$\chi^2/\nu > 1$ (either the input-data uncertainties have been underestimated
or the theory is bad): there is a third possibility, namely that one may encounter an improbable statistical fluctuation,~\cite{Gregory} whereby the computed value of $\chi^2/\nu$ is large (especially
if $\nu$ is small-ish),
even though the theory is correct. (Even more troublingly, you may find 
$\chi^2/\nu \approx 1$ even though the theory is bad.) Simply put, improbable things do happen;
Richard Feynman phrased this as follows:\cite{Feynman}
``You know, the most amazing thing happened to me tonight. I was coming here, on the way to the lecture, and I came in through the parking lot. And you won't believe what happened. I saw a car with the license plate ARW 357. Can you imagine? Of all the millions of license plates in the state, what was the chance that I would see that particular one tonight? Amazing!''


\begin{figure*}[t]
\centering
   \begin{subfigure}{0.49\textwidth} \centering
     \includegraphics[scale=0.45]{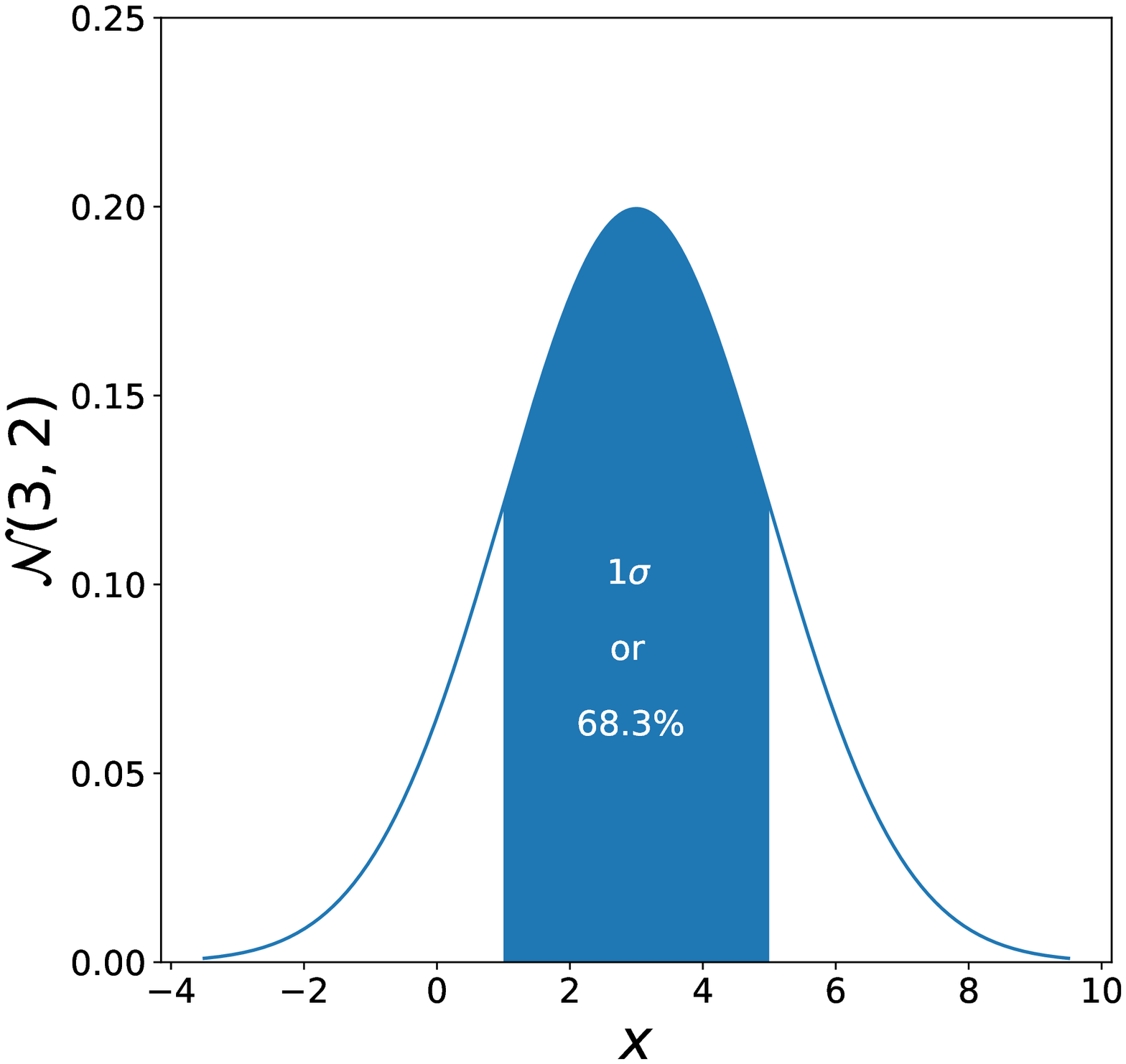}
     \caption{}
   \end{subfigure}
   \begin{subfigure}{0.49\textwidth} \centering
     \includegraphics[scale=0.45]{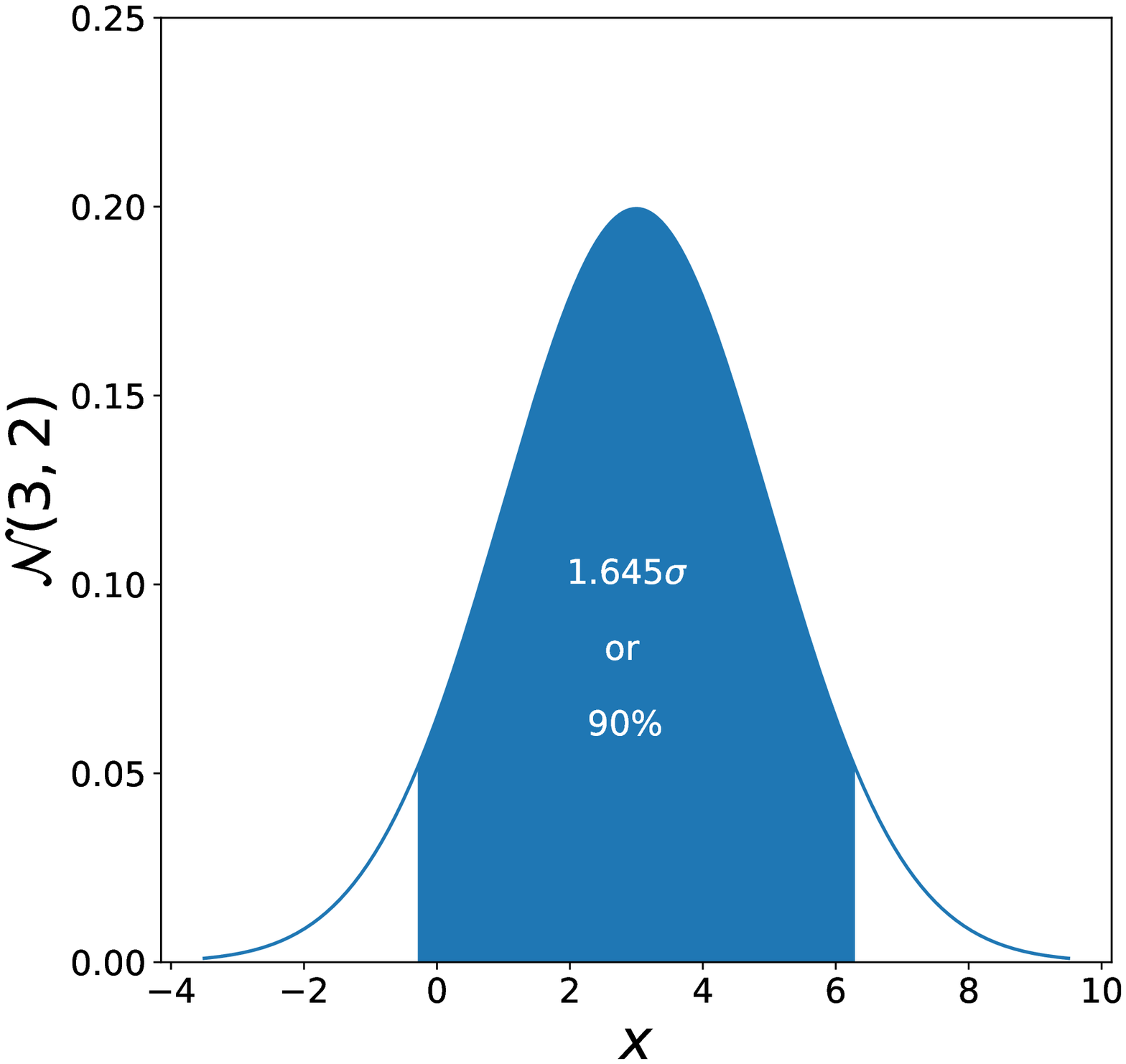}
     \caption{}
   \end{subfigure}
\caption{One-dimensional normal distribution, also visualizing
the percentage of values inside a given range: one standard deviation (left)
and 1.645 standard deviations (right).} \label{fig:empirical}
\end{figure*}

Things are even worse when it comes to model selection:\cite{Andrae} 
our argument in Eq.~(\ref{eq:Bayes7}) relied on our having access to
the true theory with the true parameter values, so we could employ
the noise term from Eq.~(\ref{eq:Bayes1}). When you don't know
those (i.e., nearly always) the numerator in the exponent 
in Eq.~(\ref{eq:Bayes6}) is not the error term 
${\cal E}_j$, so you cannot follow the derivation from standard
normal random variables to the chi-squared distribution of
Eq.~(\ref{eq:chisqdistrib0}). (Incidentally, 
if you truly did know the true theory and the true parameter values, you wouldn't
be interested in model selection in the first place.)
This means that you \textit{cannot} quantify the
spread in $\chi^2$ values by using the variance 
of $2\nu$ corresponding to Eq.~(\ref{eq:chisqdistrib0}).
Without going into more detail, we note in passing that in frequentist inference (where, e.g., employing a Bayes factor\cite{Robert} is not an option)
there are alternative techniques available that can help with model selection, e.g.,
cross-validation or bootstrapping. 
Putting it all together:

{\bf Correction \#2} \textit{You should avoid over-relying on single-number summaries of 
complex datasets; the $\chi^2$ statistic is no exception to
this general point. Blindly aiming for a $\chi^2/\nu$ of $1$ (e.g., 
over-worrying about underestimated input-data uncertainties)
might lead you to ignore
rare (but possible) statistical fluctuations. Similarly, $\chi^2$-based model selection is problematic, as the statistical foundation of thinking in terms of the chi-squared
distribution is invalid. Statisticians have developed several alternative tools
that help one to assess the goodness-of-fit or to do model selection.}

\subsection{Confidence intervals for model parameters}
\label{sec:CI}

Sections~\ref{sec:MLE}~and~\ref{sec:chisq} focused on different aspects
of essentially the same task: parameter estimation given a dataset and a 
model form. Of course, just as important (if not more important) is the question
of the uncertainty associated with the point estimates provided by 
likelihood maximization/$\chi^2$ minimization. 
The $\hat{\mathbf{c}}$ of Eq.~(\ref{eq:ls18_proba}) is  
a statistic (i.e., a function of the data $\mathbf{Y}$)
so it inherits the randomness of the data: this is why 
Eq.~(\ref{eq:ls18_proba}) also includes an expression for the variance
in each parameter estimate, $\hat{\bm{\Sigma}}_{kk}$. 
Qualitatively, if you have a huge dataset whose data points have tiny uncertainties,
and your theory captures the overall trend very well, then you would expect
that your parameter estimates have small associated uncertainties.
The meaning and
relevance of these parameter uncertainties are very often mishandled
in the literature, bringing us to:

{\bf Mistake \#3} ``\textit{Confidence intervals---probabilistic bounds on the errors in the parameters. Confidence intervals are determined from the standard errors in the parameters together with a desired probability that the bounds are correct. Under the assumptions [that the model errors are independent and normally distributed], the true value of the parameter will lie within the interval with, say, 90\% or 95\% probability.}''

Let's immediately get something out of the way: 
the modifier \textit{confidence} interval
is clearly a misnomer: one's confidence is typically \textit{low} when
the interval is big. Other terms have been proposed in the literature (e.g., uncertainty interval
or compatibility interval) but have not been widely adopted. Turning now to the 
interpretation of a confidence interval: in one dimension, the idea is motivated
by what is known as the \textit{empirical rule} 
for the normal distribution (which will re-appear below). As you can see in the left panel of Fig.~\ref{fig:empirical},
for the density function ${\cal N}(\mu, \sigma)$ of Eq.~(\ref{eq:Bayes3})
the fractional area in the interval $(\mu - \sigma, \mu + \sigma)$ is approximately $0.683$, i.e., $68.3\%$. Similarly, 
the fractional area in $(\mu - 2\sigma, \mu + 2\sigma)$ is 
$95.4\%$ and that in $(\mu - 3\sigma, \mu + 3\sigma)$ is 
$99.7\%$. This is why the term ``empirical rule'' is sometimes used
interchangeably with the term ``$68.3-95.4-99.7$ rule'' (see, however, section~\ref{sec:empirical}). As the right panel of Fig.~\ref{fig:empirical}
shows, you don't have to limit yourself to integer multiples of the standard deviation:
for example, the fractional area in $(\mu - 1.645\sigma, \mu + 1.645\sigma)$ is 
roughly $90\%$ and the fractional area in $(\mu - 1.96\sigma, \mu + 1.96\sigma)$ is 
roughly $95\%$.

The previous paragraph, and Fig.~\ref{fig:empirical},
corresponded to a simple Gaussian probability density ${\cal N}(\mu, \sigma)$, i.e.,
had nothing to do with parameter estimation. The intricacy arises (as in the 
quote above) when one tries to use the language of confidence intervals.
To see what is wrong with the (quite popular) Mistake \#3, let us (once again) turn to 
the two datasets in Fig.~\ref{fig:rawdata} for which, as you may recall,
the true parameter value for the first parameter was 
 $c_{\star, 0} = 2$. Applying the mentality of our quote would imply
 that there is a $68.3\%$ probability that the interval $(\hat{c}_0 - \sqrt{\hat{\Sigma}_{00}}, \hat{c}_0 + \sqrt{\hat{\Sigma}_{00}})$ contains the true parameter
 value $c_{\star, 0} = 2$. 
(To streamline the presentation, we are here ignoring the fact that we are faced 
with an ellipse (not a simple interval), given that $\hat{\bm{\Sigma}}$ is in general
not diagonal---see section~\ref{sec:empirical}.) 
 But what exactly does that mean? 
 Applying this prescription to the dataset (and MLE evaluation) of
 the left panel, we find the interval 
$(1.807, 2.117)$, whereas for the right panel we find the interval 
$(1.681, 1.966)$. What does it mean to say that there is a 
$68.3\%$ probability that the interval $(1.807, 2.117)$ contains the number 2?
This is nonsensical: the interval $(1.807, 2.117)$ clearly contains the number 2
(and if you choose to employ the language of probability, then 
the corresponding probability is obviously 100\%). Similarly, the interval 
$(1.681, 1.966)$ clearly does not contain the number 2, with a probability
of 100\% (i.e., contains the number 2 with a probability of 0\%).
In both cases, the relative fraction $68.3\%$ seems to be totally irrelevant. 

\begin{figure}[t]
\centering
\includegraphics[width=0.49\textwidth]{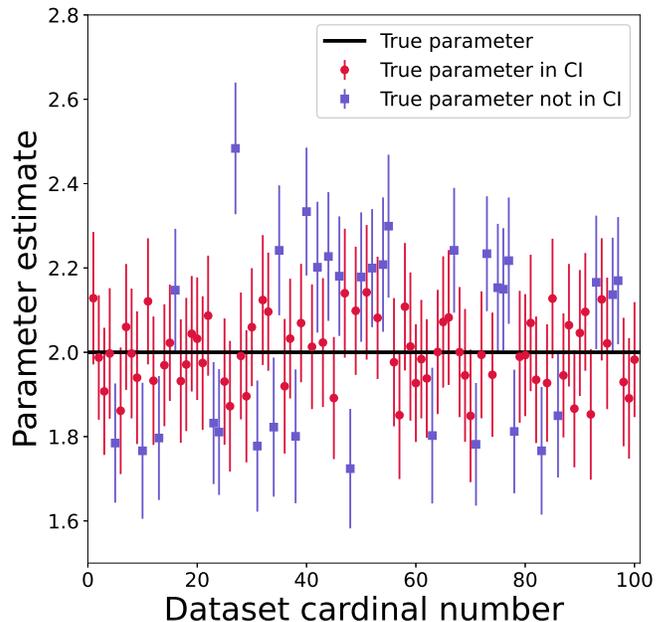}
\caption{Confidence intervals (CI) for the parameter $c_0$, computed for 100 synthetic
datasets.} 
\label{fig:confidence}
\end{figure}

At its core, Mistake \#3 arises from a fundamental misunderstanding of the meaning
of \textit{probability}: in frequentist statistics, probability refers to the
long-run relative frequency of occurrences. Symbolically, our confidence interval gives
rise to the equation:
\begin{equation}
\mathbb{P} \left ( \hat{c}_0 - \sqrt{\hat{\Sigma}_{00}} \leq c_{\star, 0} \leq \hat{c}_0 + \sqrt{\hat{\Sigma}_{00}} \right ) \approx 0.683
\label{eq:ci68}
\end{equation}
For a single interval, e.g., 
$(1.681, 1.966)$ there is no ``long'' run involved. 
Instead, Eq.~(\ref{eq:ci68}) should be interpreted as follows: 
since what is random is the interval (not the true parameter), 
to compute the probability of Eq.~(\ref{eq:ci68}) you need many different
confidence-interval evaluations, each corresponding to a different dataset.  
We have done this in Fig.~\ref{fig:confidence}, for which we generated 100 datasets
(from the same underlying function, adding noise terms of the same
form but different value each time) and the corresponding 100 confidence
intervals for the parameter $c_0$. 
For this specific case, 67 out of 100 confidence intervals contain the true parameter value; we are dealing with a $1\sigma$ random interval, for which we expect a \textit{coverage} of 68.3\%. (Reader, these are ``typical'' results.) Similarly, for $2\sigma$ confidence intervals we find the proportion of those that contain the true parameter value to be 98 out of 100   
(and for $3\sigma$ confidence intervals it is 99 out of 100).
Simply put, it is wrong to take a single one of these confidence intervals 
and try to interpret it as ``mostly'' containing the true parameter value:
for a frequentist interpretation of probability, you need a long run.

A fine-grained analysis of this misconception is carried out in Ref.~\citenumns{Morey}, 
which is characteristically titled ``The fallacy of placing confidence in confidence intervals''. The authors of that reference distinguish between 
three interrelated issues: (a) the fundamental confidence fallacy (i.e.,
making the leap from a 68.3\% probability that a random interval contains
the true value to a 68.3\% probability that a given observed interval 
contains the true value), (b) the precision fallacy (i.e.,
that a confidence interval's width tells us how precisely we know the parameter), 
and (c) the likelihood fallacy (i.e., that a confidence interval indicates the 
likely values of the parameter). All three of these very widespread misunderstandings follow from
a folk understanding of confidence intervals. 

Before closing this subsection, we note that one \textit{can} arrive at an interval
that is interpreted as per the ``folk expectation'': in a Bayesian framework, 
where the likelihood is folded together with the prior to produce the posterior
distribution, one may generate a \textit{credible interval} with the desired interpretation
(i.e., a 68.3\% rational degree of belief 
given the single dataset at our disposal).
Of course, as with everything in life, you win some you lose some: the credible
interval that is intuitively easy/natural to interpret, will not in general
have the same frequency coverage properties, i.e., after generating many datasets
and the corresponding credible intervals, it will not be true that 
68.3\% of the credible intervals will contain the true parameter. 
In short:

{\bf Correction \#3} \textit{A single confidence interval should never be interpreted
in terms of the probability that it contains the true parameter value. Since the true parameter value is fixed (though unknown), a given confidence interval either contains
the true parameter or it doesn't. The (frequentist) concept of a confidence interval
coverage takes on a probabilistic interpretation when you consider many experiments,
i.e., distinct datasets giving rise to distinct confidence intervals ($Q\%$ of which
will contain the true parameter value). If you really want to interpret a given
interval in terms of the probability that it contains the true parameter, you should
be using a (Bayesian) credible interval.}

\begin{figure*}[t]
\centering
   \begin{subfigure}{0.49\textwidth} \centering
     \includegraphics[scale=0.45]{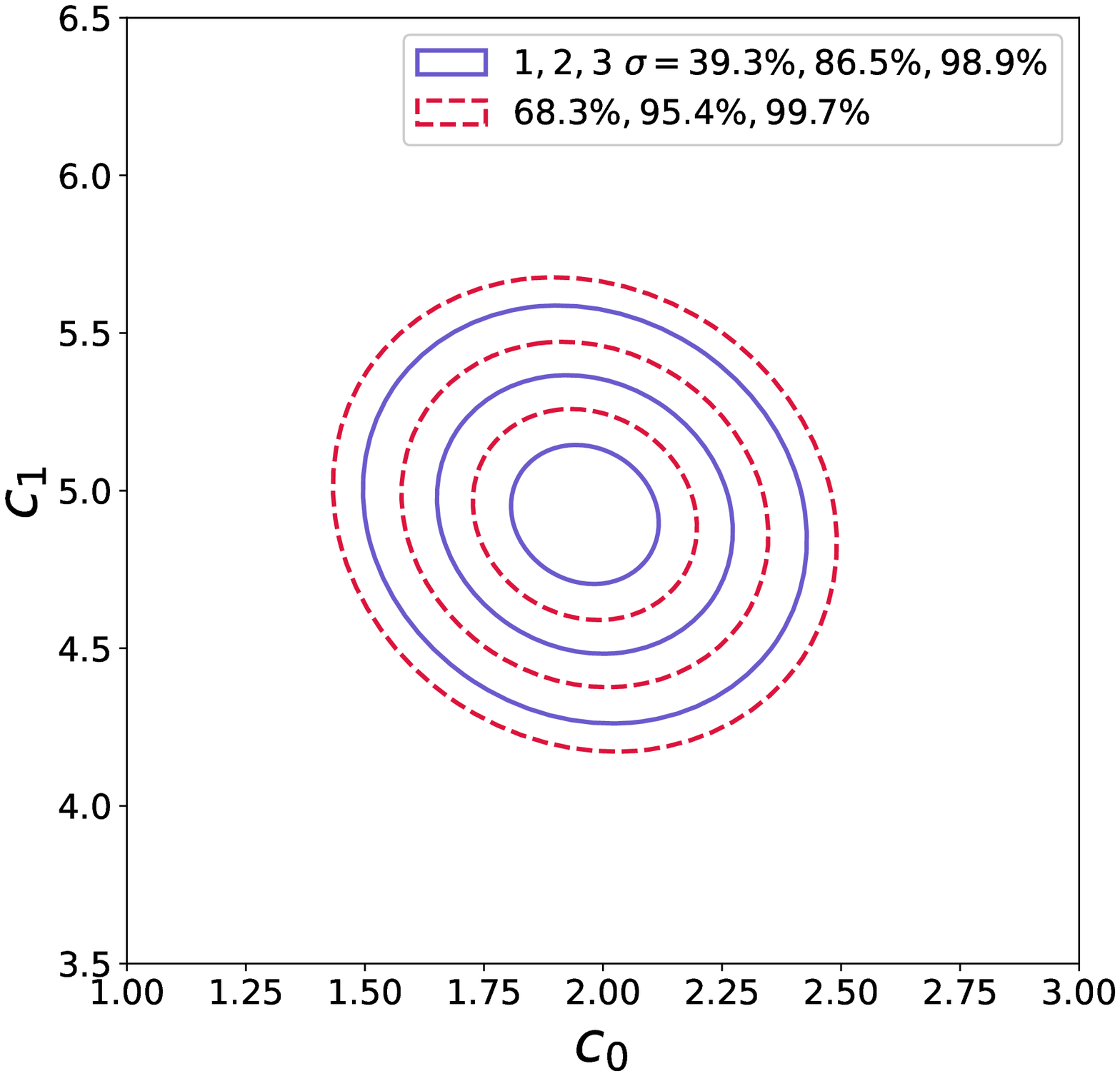}
     \caption{}
   \end{subfigure}
   \begin{subfigure}{0.49\textwidth} \centering
     \includegraphics[scale=0.45]{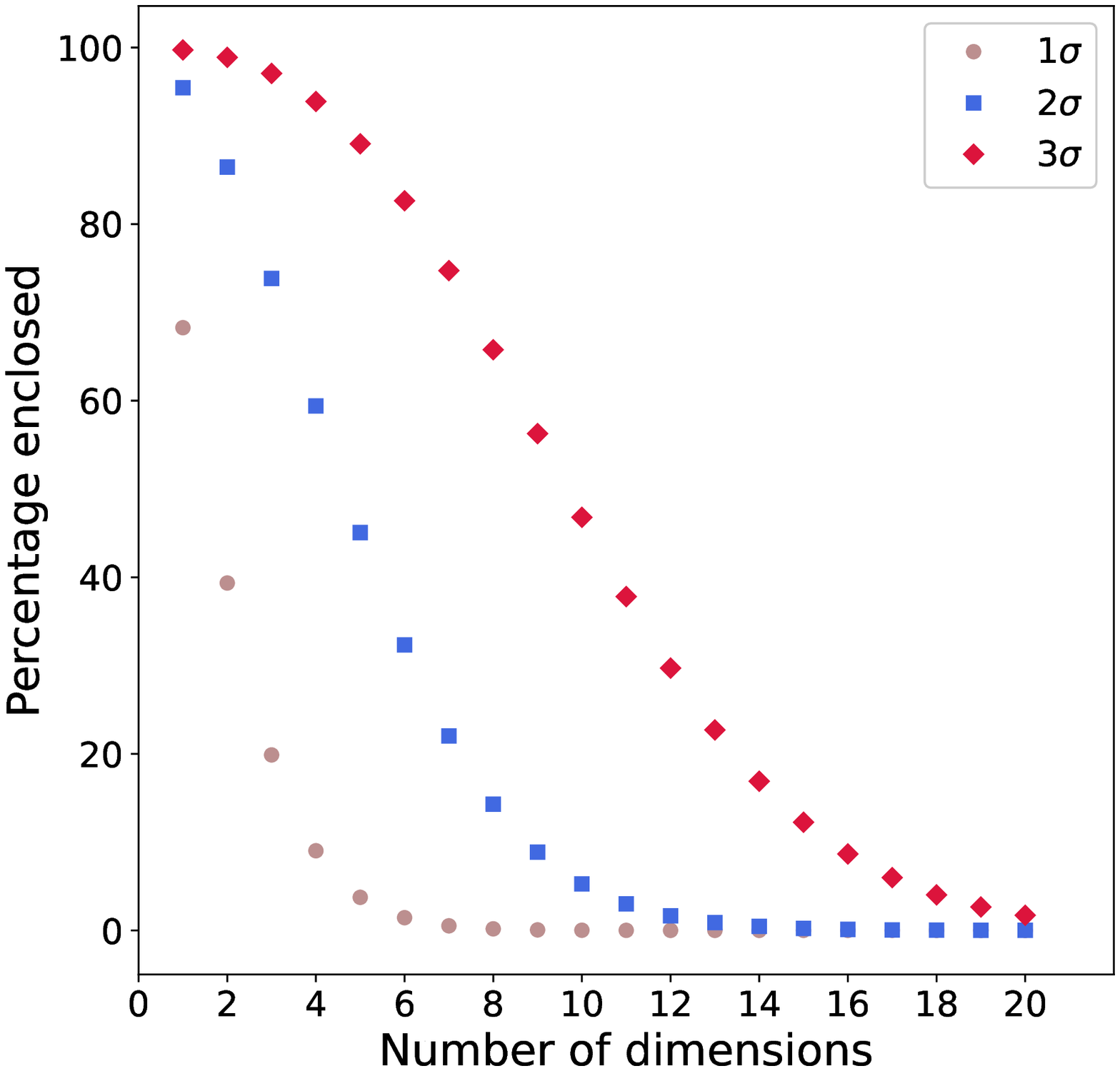}
     \caption{}
   \end{subfigure}
\caption{Confidence ellipses for a two-dimensional normal distribution (left)
and the many-dimensional version of the $1, 2, 3~\sigma$ rule (right).} \label{fig:ellipse}
\end{figure*}

\subsection{Empirical rule in the multivariate case}
\label{sec:empirical}

The previous subsection addressed the problem of constructing a confidence
interval for a single model parameter. Of course, in practical applications
we are typically faced with multivariate problems, in which case we have to
construct \textit{confidence regions} (sometimes known as
\textit{confidence sets}). Human intuition doesn't work very well in many 
dimensions, but we are fortunate that our running example from Fig.~\ref{fig:rawdata}
involves only two parameters, in which case visualizing things is still a possibility.
The next mistake in our list relates to the afore-mentioned empirical rule, namely
the rule that tells us what the probability enclosed in a given number of standard
deviations is. Even if you have mastered Correction \#3, namely the meaning of
the term probability in this context, perusing the introductory literature (or even research works) will
convince you that there is still room for a further misunderstanding:

{\bf Mistake \#4} ``\textit{The result of the measurement can be visualized as iso-density contours in the parameter space. These contours are elliptical in the case of Gaussian likelihood functions, and are therefore called ``error ellipses''. Typically one is interested in the contour that encloses 68.3\% of the likelihood, corresponding to one Gaussian deviation or ``one sigma'', or 95.4\% corresponding to two Gaussian standard deviations or ``two sigma''.}''

Basically, the claim here is that the empirical rule is equivalent to the 
$68.3-95.4-99.7$ rule, even in more than one dimensions. To understand why this is wrong,
let us turn to our expression in Eq.~(\ref{eq:ci68}): it is hard to see how to generalize
that to more dimensions, so let us first recast it into the form:
\begin{equation}
\mathbb{P} \left ( \left | \frac{\hat{c}_0 - c_{\star, 0}}{\sqrt{\hat{\Sigma}_{00}}} \right | \leq 1 \right ) \approx 0.683
\label{eq:ci68_2}
\end{equation} 
where we would have $|(\hat{c}_0 - c_{\star, 0}) / \sqrt{\hat{\Sigma}_{00}}| \leq 2$ for the $2 \sigma$ interval, and so on. The fraction
inside the absolute value here takes 
the form $|(X - \mu)/ \sigma|$, where $X$ is a random variable. (This is a rescaling operation that is tremendously
common when dealing with Gaussian distributions.) 
You may recall that in section~\ref{sec:CI} we, somewhat cryptically, referred
to the fact that $\hat{\bm{\Sigma}}$ is in general
not diagonal, implying that Eq.~(\ref{eq:ci68}), and therefore also Eq.~(\ref{eq:ci68_2})
doesn't tell us the full story; it is now time to see what this means. 
First, we recall from Eq.~(\ref{eq:ls18_proba}) that $\hat{\mathbf{c}}$ is a linear
transformation of $\mathbf{Y}$; since we know from Eq.~(\ref{eq:Bayes6}) that
$\mathbf{Y}$ obeys a Gaussian distribution, then so will $\hat{\mathbf{c}}$
($N$ and $n$ dimensional, respectively). 
As mentioned in section~\ref{sec:MLE}, the MLE estimator is unbiased, i.e., 
$\mathbb{E}(\hat{\mathbf{c}}) = \mathbf{c}_{\star}$. Combined with the covariance
matrix for our parameter estimates, again from Eq.~(\ref{eq:ls18_proba}), 
we have thereby seen that $\hat{\mathbf{c}}$ obeys the distribution:
\begin{align}
&P(\hat{\mathbf{c}}; \mathbf{c}_{\star}, \hat{\bm{\Sigma}}) 
= {\cal N}\left ( \mathbf{c}_{\star}, \hat{\bm{\Sigma}} \right ) \nonumber \\
&= \frac{1}{(2\pi)^{n/2} |\hat{\bm{\Sigma}}|^{1/2}} \exp 
\left [ - \frac{1}{2} \left (\hat{\mathbf{c}} - \mathbf{c}_{\star} \right )^T \hat{\bm{\Sigma}} ^{-1} \left (\hat{\mathbf{c}} - \mathbf{c}_{\star} \right ) \right ]
\label{eq:Baynew2}
\end{align}
Observe that this is our first multivariate Gaussian: Eq.~(\ref{eq:Bayes6})
and Eq.~(\ref{eq:parent}) were written as products of univariate Gaussians.
The form of Eq.~(\ref{eq:Baynew2}) is more general, i.e., the prefactor involving
the determinant also works when $\hat{\bm{\Sigma}}$ is not diagonal.

Our result in Eq.~(\ref{eq:ci68_2}) was implicitly assuming a univariate Gaussian; 
since we know that $\hat{c}_0$ is not an independent parameter, i.e., it is 
part of the random vector $\hat{\mathbf{c}}$, let us now see how to generalize
our statements on probability. 
The natural generalization of $|(\hat{c}_0 - c_{\star, 0}) / \sqrt{\hat{\Sigma}_{00}}|$ to many dimensions is 
via the \textit{Mahalanobis distance}, namely:
$\sqrt{ \left (\hat{\mathbf{c}} - \mathbf{c}_{\star} \right )^T \hat{\bm{\Sigma}}^{-1} \left ( \hat{\mathbf{c}} - \mathbf{c}_{\star} \right )}$. The analogue
of Eq.~(\ref{eq:ci68_2}) would now be to check if the Mahalanobis distance
is less than, say, 1 and thereby produce a $1\sigma$ interval. For example, a simple test with the 100 synthetic datasets used in 
Fig.~\ref{fig:confidence} leads to $36, 88, 98$ Mahalanobis distances
enclosed within $1, 2, 3~\sigma$, respectively. Of course, these are not
asymptotic results (since they correspond to only 100 different 
instances of $\hat{\mathbf{c}}$) but, still, you can already see that,
for this two-dimensional problem, $1\sigma$ corresponds to roughly
40\% \textit{contra} what the quote in Mistake \#4 claimed.

Let's try to be a bit more systematic:\cite{Bajorski} we already know from Eq.~(\ref{eq:Baynew2})
that  $\hat{\mathbf{c}}$ obeys a normal distribution; this means that 
$\hat{\bm{\Sigma}} ^{-1/2} (\hat{\mathbf{c}} - \mathbf{c}_{\star} )$
obeys a standard normal distribution. But then we see that the argument of the 
square root in the Mahalanobis distance definition, $\left (\hat{\mathbf{c}} - \mathbf{c}_{\star} \right )^T \hat{\bm{\Sigma}}^{-1} \left ( \hat{\mathbf{c}} - \mathbf{c}_{\star} \right )$, is simply the square of a (multivariate) standard normal distribution so,
as per the same lemma we employed in section~\ref{sec:chisq}, 
it will obey an $n$-dimensional---not $\nu$-dimensional as in Eq.~(\ref{eq:chisqdistrib0})---chi-squared distribution. We have thereby shown that:
\begin{align}
&\mathbb{P} \left ( \sqrt{ \left (\hat{\mathbf{c}} - \mathbf{c}_{\star} \right )^T \hat{\bm{\Sigma}}^{-1} \left ( \hat{\mathbf{c}} - \mathbf{c}_{\star} \right ) } \leq p \right ) \nonumber \\
&= \mathbb{P} \left ( \left (\hat{\mathbf{c}} - \mathbf{c}_{\star} \right )^T \hat{\bm{\Sigma}}^{-1} \left ( \hat{\mathbf{c}} - \mathbf{c}_{\star} \right )  \leq p^2 \right ) = F_{n}(p^2)
\label{eq:ci99}
\end{align} 
In the first equality we simply eliminated the square root. In the second equality
we made use of $F_{n}$, the cumulative distribution function corresponding to the
chi-squared distribution.

We are now ready to produce some numerical values. In one dimension, 
Eq.~(\ref{eq:ci99}) leads to a confidence interval, for which the 
relative fractions enclosed when $p = 1, 2, 3$ are 
$68.3\%,~95.4\%,~99.7\%$ respectively; both here and below, the 
meaning of ``relative fraction'' or of ``probability'' enclosed 
is as per section~\ref{sec:CI}, i.e., in the long run. In two dimensions 
we are faced with a confidence ellipse and the 
relative fractions enclosed when $p = 1, 2, 3$ are 
$39.3\%,~86.5\%,~98.9\%$ respectively; these are visualized with solid curves in the 
left panel of Fig.~\ref{fig:ellipse} for the dataset from the left panel of
Fig.~\ref{fig:rawdata}. 
Since in practice we don't know the $\mathbf{c}_{\star}$ in Eq.~(\ref{eq:ci99}),
we get ellipses centered at the $\hat{\mathbf{c}}$ corresponding to a given dataset 
$\mathbf{y}$; one of the confidence ellipses here is analogous to a single confidence interval (i.e., 
a point with an associated error bar) from Fig.~\ref{fig:confidence}.
Of course, there is nothing making
us employ integer values of $p$: we could, instead, choose to work
with a specified value of $F_{n}(p^2)$ (e.g., $0.683$) and then
reverse-engineer the corresponding value of $p$. We have done this
in the left panel of Fig.~\ref{fig:ellipse}, where we also
show (with dashed curves) 
the confidence ellipses corresponding to $68.3\%,~95.4\%,~99.7\%$.
Crucially, for this two-dimensional problem these are \textit{different} from the 
confidence ellipses corresponding to $p = 1, 2, 3$; most obviously, 
a $1\sigma$ ellipse in two dimensions encloses a considerably smaller relative fraction
than the $68.3\%$ ellipse does.

We can keep playing this game: 
in three dimensions we will be faced with an ellipsoid
and in more dimensions with a hyper-ellipsoid. As you can see in the right
panel of Fig.~\ref{fig:ellipse} (this time applying the right-hand side of Eq.~(\ref{eq:ci99}), i.e., not limiting ourselves to a given dataset), as the dimensionality grows 
the corresponding $1, 2, 3~\sigma$ confidence regions enclose a considerably
smaller relative fraction. This is most noticeable for the $1\sigma$ case,
where the percentage enclosed rapidly drops as $n$ is increased; even for the 
$3\sigma$ case, however, the relative fraction enclosed for, say, $n=10$ is only $47\%$.

{\bf Correction \#4} \textit{The empirical rule takes the form 
$68.3\%,~95.4\%,~99.7\%$ only for a single variable (i.e., in a one-parameter problem).
The Mahalanobis distance provides us with a natural way to generalize to the many-dimensional case. This leads to confidence ellipses enclosing $39.3\%,~86.5\%,~98.9\%$
of the probability in two dimensions. In higher-dimensional problems the relative
fraction enclosed in the $1, 2, 3~\sigma$ confidence regions is increasingly smaller
or (to reverse the argument) you would need to employ a $p \sigma$ interval
where $p$ is increasingly large if you wish to enclose, say, $95.4\%$ of the probability.
}

\subsection{What is random in frequentist vs Bayesian regression}
\label{sec:bayes}

As noted above, most of our discussion up to this point has been (sometimes implicitly)
focused on frequentist approaches to linear regression; this is merely a consequence
of the fact that the relevant textbooks with a scientific or engineering 
readership are still 
largely unBayesian in their outlook (sometimes even going so far as to take
cheap shots at those foolish enough to think differently); despite the well-known inertia 
exhibited by physical scientists (leading to FORTRAN 77 being used as the implementation language in books written 55 years later), this is, thankfully, now beginning to change. On the other hand, machine-learning 
textbooks are overwhelmingly Bayesian in their viewpoint; more often than not,
such works remind one of so many histories of philosophy, hurriedly dispatching
alternative viewpoints, in the process throwing the methodological principle of charity out the window. 

Obviously, we cannot discuss philosophical/conceptual issues in depth in the present
article, 
so let
us, instead, briefly outline the different approaches to probability; on this topic, the reader may
enjoy Ref.~\citenumns{Diaconis}, a work which is addressed  
at (almost) a layperson. In
the frequentist viewpoint, as we've already seen, probability represents the long-run relative frequency
of occurrences; for example, after a large number of coin flips, a table shows
us that roughly half were heads and half tails. In the Bayesian viewpoint, 
probability represents a rational degree of belief about something given 
current knowledge; for example, we expect that for a fair coin the next flip
is just as likely to give heads as tails. Almost everyone is in agreement on 
such summaries; things get a bit iffy when we turn to the interpretation of regression:

{\bf Mistake \#5} ``\textit{In the frequentist approach, $c$ is treated as an unknown fixed constant, and the data is treated as random. In the Bayesian approach, we treat the data as fixed (since it is known) and the parameter as random (since it is unknown).}''

There is quite a bit to unpack here. Let us start from what we touched upon 
more often above, namely the frequentist approach: it is true 
that a frequentist would take the data-generating
parameters $\mathbf{c}_{\star}$ appearing in Eq.~(\ref{eq:Bayes1})
to be constant. (Of course, it's not clear what the quote above meant by emphasizing
that the parameter is unknown \textit{in the Bayesian approach}, since 
$\mathbf{c}_{\star}$ is unknown in the frequentist approach, as well.)
That being said, the main set of parameters appearing in a frequentist approach is
not $\mathbf{c}_{\star}$ but the $\hat{\mathbf{c}}$ of Eq.~(\ref{eq:ls18_proba}), namely
the result of thinking in terms of a general likelihood 
$L(\mathbf{c}) = P(\mathbf{y} ; \mathbf{c}, \bm{\Phi}, \bm{\Sigma}_d)$
and then finding the argument which maximizes that. But, as Eq.~(\ref{eq:ls18_proba}) clearly
shows, $\hat{\mathbf{c}}$ is a function of the data and is therefore a random vector.
That being said, even in a frequentist approach you are typically faced with only
a single dataset $\mathbf{y}$ (so you produce a single $\hat{\mathbf{c}}$ point estimate). 
The dataset $\mathbf{y}$ is, indeed, treated as random in the sense of 
being a realization of the random vector $\mathbf{Y}$ (but that is relevant only when one considers, e.g., the interpretation of the covariance matrix for our parameter estimates, $\hat{\bm{\Sigma}}$). In short, in so far as the frequentist approach is concerned,
you could interpret the above quote to be correct, but you could also read it as 
being completely wrong. 

Turning now to the Bayesian viewpoint, let's start from the (easier) point regarding
the data: while it's true that we wrote down Bayes' rule in Eq.~(\ref{eq:Bayesrule3})
for a single dataset $\mathbf{y}$, that same equation contains the likelihood 
$P(\mathbf{y} | \mathbf{c}; \bm{\Phi}, \bm{\Sigma}_d)$ in the numerator, an entity
which most clearly shows that the dataset $\mathbf{y}$ is a realization of a general
distribution for $\mathbf{Y}$. In other words, while you focus on a single dataset
(as afore-mentioned, just as you did in the frequentist approach to get the point estimate $\hat{\mathbf{c}}$), you are also dealing with a statistical model reflecting what you know about how a given $\mathbf{c}$
gives rise to $\mathbf{y}$'s.
Finally, while we noted in section~\ref{sec:CI} that (frequentist) confidence intervals
will have the expected coverage (across alternative datasets \textit{in the absence of systematic errors}), and (Bayesian) 
credible intervals will in general not exhibit that coverage, there is nothing keeping
you from examining the frequentist coverage (across alternative datasets) of your Bayesian theory; as a matter of fact, this is something that is routinely done 
in Bayesian approaches, as reflected already in chapter 1 of Ref.~\citenumns{Gelman}.

Still on the subject of the Bayesian approach to regression, we now turn to the
(thornier) question of the status of the parameters. Somewhat frustratingly,
many Bayesian expert treatments are tight-lipped on the matter of 
the underlying function/the true parameters $\mathbf{c}_{\star}$/the 
data-generating distribution.
We believe there are two reasons for this: first, when studying other
fields it is often awkward to be thinking in terms of ``true fixed parameters''
(consider, e.g., the study of economic growth). Things are considerably
different in physical science, 
where nearly all physicists would agree that, e.g., the mass
of the electron is a constant that we would be thrilled to be able to compute from
first principles---and, absent that, can try to extract from experimental 
measurements.
(Of course, even in physics, 
advocates of Quantum Bayesianism would advocate against thinking 
in terms of ``elements of physical reality''.) \cite{Fuchs} 
Second, a careful study of the foundations of the subject will expose you 
to an (orthodox) viewpoint\cite{BernardoSmith,Sprenger} which is 
\textit{subjectivist} and \textit{operational}, namely one in which the focus is on individual beliefs about
how to make decisions (i.e., not on the true parameters). In lieu of entering that discussion
ourselves, we simply reach for an \textit{argumentum ab auctoritate}, quoting
two prominent Bayesians on the subject: ``Notice that (in sharp contrast to conventional statistics) \textit{parameters are treated as random variables} within the Bayesian paradigm. This is not a description of
their variability (parameters are typically \textit{fixed unknown} quantities) but a description of the \textit{uncertainty} about their true values.'' (J. M. 
Bernardo)~\cite{Bernardo} and ``This would be another form of the mind projection fallacy, confusing reality with a state of knowledge about reality. In the problem we are discussing, $c_{\star}$ is simply an unknown constant parameter; what is distributed is not the parameter, but the probability.'' (E. T. Jaynes).~\cite{Jaynes} 

Obviously, quoting from the works of acknowledged experts does not rise to the level of a sustained argument, but this should suffice to show that the (extremely widespread) summarization in Mistake \#5 is 
certainly oversimplified and arguably flat-out incorrect. 
There are several corollaries to this haphazard treatment of the conceptual
underpinnings of frequentist vs Bayesian approaches to regression; for example, one often hears mention of the ``dataset likelihood'',
though the term ``likelihood'' should be more properly reserved for the 
parameters. Similarly, we broke with the overwhelming majority of the literature
by not using $|$ in the $P(\mathbf{y} ; \mathbf{c}, \bm{\Phi}, \bm{\Sigma}_d)$
of Eq.~(\ref{eq:Bayes6}), since $\mathbf{c}$ is not (the realization of) a random vector there. (Compare
with the Bayesian setting of Eq.~(\ref{eq:Bayesrule3}), which \textit{does} employ $|$.)
Even so, these questions are not as important as the central issue discussed above,
namely what is random (and what is not) in frequentist vs Bayesian approaches to regression. It's fair to say that this 
is considerably more involved than introductory treatments let on. (Again, this is 
perhaps understandable, though not excusable.)

{\bf Correction \#5} \textit{It should come as no surprise that 
bumper-sticker formulations can lead one astray: there is no substitute for 
careful thinking and clear exposition. It is certainly defensible (both from
frequentist and from Bayesian viewpoints) to take the data-generating parameters $\mathbf{c}_{\star}$
to be unknown constants. Similarly, both frequentist and Bayesian approaches
can view the data as being a single realization $\mathbf{y}$ of the random vector
$\mathbf{Y}$. The real difference between the two approaches (except for the inclusion
or not of prior information) is on whether the 
parameters are taken to be a random vector or not. 
Of course, even in a frequentist setting the MLE estimator $\hat{\mathbf{c}}$ is a random
vector; the crucial difference is that expectations there are taken across datasets (i.e., in $N$ dimensions) whereas in a Bayesian approach they are across parameter values (i.e.,
in $n$ dimensions).
}

\subsection{Posterior predictive distribution, noise, and samples}

\begin{figure*}[t]
\centering
   \begin{subfigure}{0.49\textwidth} \centering
     \includegraphics[scale=0.45]{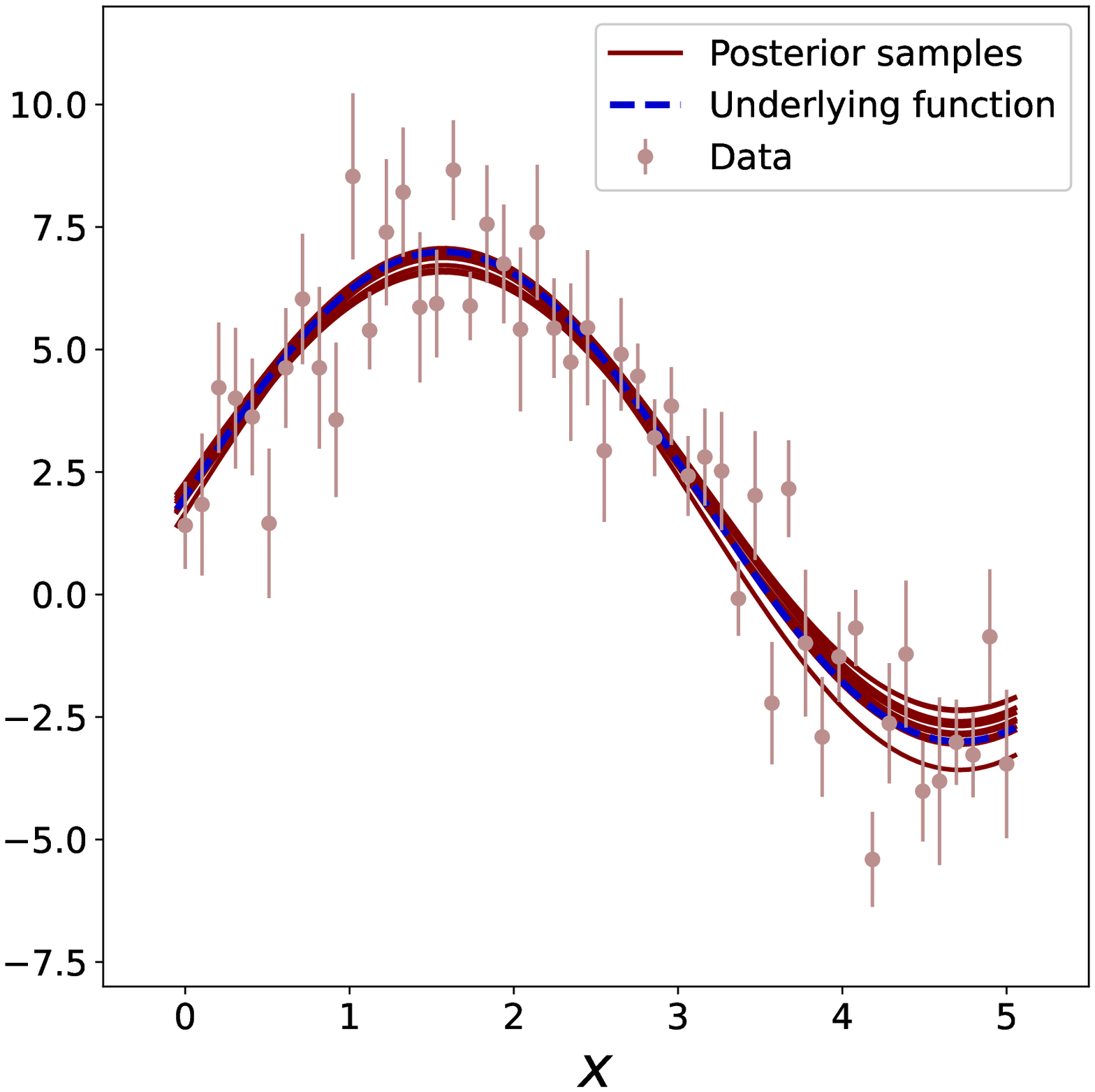}
     \caption{}
   \end{subfigure}
   \begin{subfigure}{0.49\textwidth} \centering
     \includegraphics[scale=0.45]{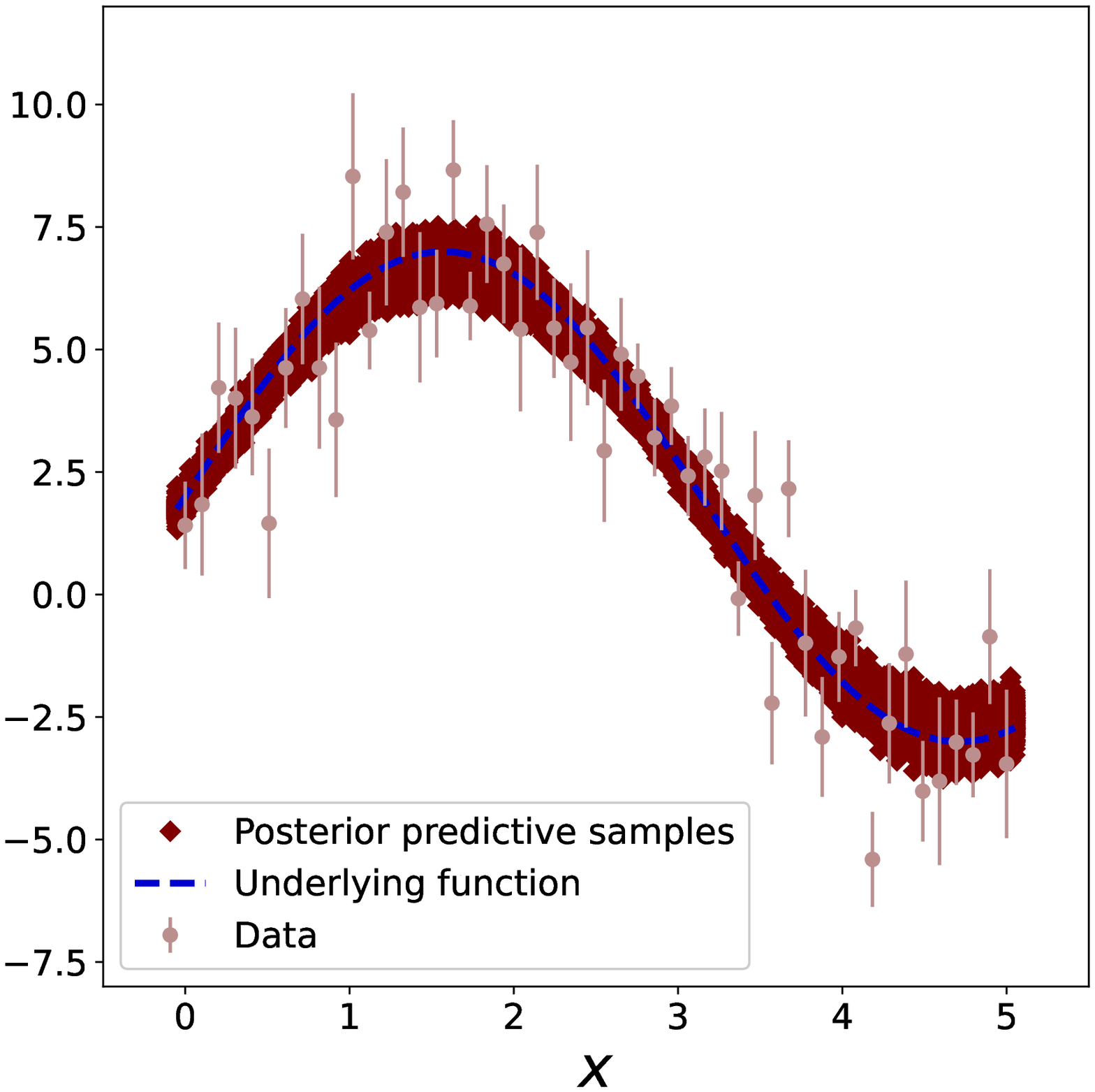}
     \caption{}
   \end{subfigure}
\caption{For the dataset from the left panel of Fig.~\ref{fig:rawdata}, 
we show predictions corresponding to samples drawn from the posterior
distribution (left) and from the posterior predictive distribution (right).} \label{fig:samples}
\end{figure*}

While section~\ref{sec:bayes} changed gears (transitioning from a frequentist
to a Bayesian viewpoint) it was quite abstract, involving no equations or figures. 
With the conceptual background now in place, we turn to more practical aspects of
what is entailed by a Bayesian approach to regression. As it so happens, our next general theme
(\textit{predictive distributions}) is actually of relevance even in a 
frequentist/MLE setting. Simply put, the idea is as follows: after one has
somehow determined the parameter values (either as a point estimate or in the form
of the posterior distribution), one would typically like to find which
\textit{predictions} correspond to that set of parameters; in the language of 
Eq.~(\ref{eq:predi1}), this means introducing a new random variable $\tilde{Y}$ starting from the (random) parameters $\mathbf{C}$:
\begin{equation}
\tilde{Y} = \sum_{k=0}^{n-1} C_k \phi_{k}(\tilde{x}) = \tilde{\bm{\phi}} \mathbf{C} 
\label{eq:predi2}
\end{equation}
To ensure this point does not get muddled in a sea of formalism, we immediately note
that this equation (modulo a change or two in the notation) is precisely of the same
form as what is given in countless discussions of least-squares fitting in numerical-methods
textbooks.
In an MLE setting the $\mathbf{C}$ in Eq.~(\ref{eq:predi2}) would be $\hat{\mathbf{c}}$
(this is precisely how the MLE predictions in Fig.~\ref{fig:rawdata} were produced); in a 
Bayesian setting it could correspond to the maximum of the posterior distribution, but it would be best
to somehow fold in the entirety of the posterior (i.e., not just a 
point estimate).
Eventually, one would like to go beyond Eq.~(\ref{eq:predi2}) by 
producing, e.g., a $3\sigma$ uncertainty
band in predicted values. Before we go too far, let us examine our sixth (and final)
quote:

{\bf Mistake \#6} ``\textit{[W]e can show that the posterior predictive distribution at a test point $\tilde{x}$ is also Gaussian:
\begin{align*}
P(\tilde{y} | \mathbf{y}; \tilde{\bm{\phi}}, \bm{\mu}_{\mathbf{c}}, \bm{\Sigma}_{\mathbf{c}}) &= 
\int d^n c~{\cal N}(\tilde{\bm{\phi}} \mathbf{c}, \sigma)~{\cal N}\left (\bm{\mu}_{\mathbf{c}}, \bm{\Sigma}_{\mathbf{c}} \right )\stepcounter{equation}\tag{\theequation}\label{eq:predibayes2}\\
&= {\cal N}(\tilde{\bm{\phi}} \bm{\mu}_{\mathbf{c}}, \overset{\huge\frown}{\sigma})
\end{align*}
where $\overset{\huge\frown}{\sigma}^2 = \sigma^2 + \tilde{\bm{\phi}}~\bm{\Sigma}_{\mathbf{c}}~\tilde{\bm{\phi}}^T $ is the variance of the posterior predictive distribution at point $\tilde{x}$ after seeing the $N$ training examples. The predicted variance depends on two terms: the variance of the observation noise, $\sigma^2$, and the variance in the parameters, $\bm{\Sigma}_{\mathbf{c}}$. 
 [... The left panel of Fig.~\ref{fig:samples} shows] 10 samples from the true posterior predictive distribution.}''
 
This mistake is made up of two sub-fallacies, one relating to noise and the other
relating to samples; let us examine them in turn. First, we discuss the meaning of 
noise; as you may recall, this is the term we used to describe the
continuous random variable ${\cal E}_j$ in Eq.~(\ref{eq:Bayes1}). 
In the present context, ``noise'' plays an additional role: in 
Eq.~(\ref{eq:predibayes2}) we are faced with an integral over a product of two
terms, both Gaussian density functions; this is a specific version of 
the product that we wrote down as 
$P(\tilde{y} | \mathbf{c}; \tilde{\bm{\phi}})  P(\mathbf{c} | \mathbf{y}; \bm{\Phi}, \bm{\Sigma}_d, \bm{\mu}_0, \bm{\Sigma}_0)$ in Eq.~(\ref{eq:predi1}). 
The second term in ${\cal N}(\tilde{\bm{\phi}} \mathbf{c}, \sigma)~{\cal N}\left (\bm{\mu}_{\mathbf{c}}, \bm{\Sigma}_{\mathbf{c}} \right )$ 
is easy enough to interpret: if you take Bayes' rule of Eq.~(\ref{eq:Bayesrule3})
and apply it for the case of a Gaussian likelihood and a Gaussian prior distribution,
the result is a Gaussian posterior distribution, with mean vector 
$\bm{\mu}_{\mathbf{c}}$ and covariance matrix $\bm{\Sigma}_{\mathbf{c}}$ as per Eq.~(\ref{eq:Baynew6}). The first term is a bit more involved: 
many sources in the literature employ it without comment, calling it a
``prediction'' or (as in the quote above) an ``observation''. 
To assume that our model for producing an $\tilde{y}$ given 
the values of $\mathbf{c}$ and $\tilde{\bm{\phi}}$, namely $P(\tilde{y} | \mathbf{c}; \tilde{\bm{\phi}})$, is Gaussian is to have ``one thought too many'' (to borrow
Bernard Williams' inimitable phrase).~\cite{Williams}
While such an assumption allows one to use a standard property of multivariate
Gaussians---thereby reaching the second line in Eq.~(\ref{eq:predibayes2})---it 
has to be juxtaposed with the inconvenient fact that we (together with 
the entire numerical-methods community) already assumed
a perfectly natural (yet non-Gaussian) model in Eq.~(\ref{eq:predi2}), namely:
\begin{equation}
P(\tilde{y} | \mathbf{c}; \tilde{\bm{\phi}}) = 
\delta(\tilde{y} - \tilde{\bm{\phi}} \mathbf{c})
\label{eq:predibayes1}
\end{equation}
To be fair, some authors make the distinction between noisy observations
and noise-free function values; 
sometimes there is even an attempt to use \textit{prediction}
for a noisy quantity and \textit{regression} for a noise-free quantity,
but other authors employ the exact opposite nomenclature; quite confusingly, a given
author often employs both (!). Here, prediction will always be noise-free, i.e.,
will take the logically true form of Eq.~(\ref{eq:predibayes1}). Let us
now see what this implies regarding
the \textit{posterior predictive distribution}:
\begin{align}
&P(\tilde{y} | \mathbf{y}; \tilde{\bm{\phi}}, \bm{\Phi}, \bm{\Sigma}_d, \bm{\mu}_0, \bm{\Sigma}_0) \nonumber \\
&\propto \int d^n c~\delta(\tilde{y} - \tilde{\bm{\phi}} \mathbf{c}) \exp 
\left [ - \frac{1}{2} \left ( \mathbf{c} - \bm{\mu}_{\mathbf{c}} \right )^T 
\bm{\Sigma}_{\mathbf{c}}^{-1}
\left ( \mathbf{c} - \bm{\mu}_{\mathbf{c}} \right )
\right ] \nonumber \\
&= {\cal N}\left ( \tilde{\bm{\phi}} \bm{\mu}_{\mathbf{c}}, \tilde{\bm{\phi}}~\bm{\Sigma}_{\mathbf{c}}~\tilde{\bm{\phi}}^T \right )
\label{eq:predi5}
\end{align}
In the first step we plugged in the (one-dimensional) Dirac delta function from 
Eq.~(\ref{eq:predibayes1}) and the Gaussian corresponding to 
Eq.~(\ref{eq:Baynew6}) into Eq.~(\ref{eq:predi1}). We do not
explicitly derive the second step, pointing instead to Ref.~\citenumns{Gezerlis}, pp. 405-407; we merely note that we used a straightforward argument involving, among other things, rotation matrices as well as the fact that Gaussians are closed under marginalization. Note that our final result is identical to that in 
Eq.~(\ref{eq:predibayes2}) if one simply sets $\sigma^2 = 0$ for the \textit{ad hoc}
observation noise employed there (as it should, given that a zero-width Gaussian
is one definition of a Dirac delta function).

Next, we turn to the second sub-fallacy, this one relating to samples. 
We first observe that the posterior predictive distribution of Eq.~(\ref{eq:predi5})
takes in a dataset (i.e.,
many $x_j$'s, $y_j$'s, and
$\sigma_j$'s), a set of basis functions
(i.e., many $\phi_k$'s), as well as a prior distribution
and a given $\tilde{x}$; it folds in the posterior distribution's
mean vector $\bm{\mu}_{\mathbf{c}}$ and covariance matrix $\bm{\Sigma}_{\mathbf{c}}$ from Eq.~(\ref{eq:Baynew6}) to tell us which $\tilde{y}$ values we should
expect.
Crucially, Eq.~(\ref{eq:predi5})
shows a \textit{univariate} Gaussian in $\tilde{y}$; 
observe that the mean $\tilde{\bm{\phi}} \bm{\mu}_{\mathbf{c}}$ and the variance
$\tilde{\bm{\phi}}~\bm{\Sigma}_{\mathbf{c}}~\tilde{\bm{\phi}}^T$ 
are \textit{numbers} (i.e., not vectors/matrices). It is therefore very strange
to read in Mistake \#6 that samples drawn from the posterior predictive distribution look
like the left panel of Fig.~\ref{fig:samples}:
shouldn't samples from a univariate Gaussian have a ``jittery'' look?
As it so happens, if you do produce 100 samples from the 
posterior predictive distribution of Eq.~(\ref{eq:predi5}) you will find 
the, decidedly non-smooth, results shown in the right panel of
Fig.~\ref{fig:samples}. 
Note that this is not a simple typo (e.g., a confusion of left vs right panel):
we have not actually encountered (nary a reference to) a plot like that in the right panel of
Fig.~\ref{fig:samples} in the literature
(outside Ref.~\citenumns{Gezerlis}). 
As it so happens, the smooth results shown in the left panel of
Fig.~\ref{fig:samples} (which frequently appear in textbooks) 
were produced in a totally different way: they result from taking samples
of the $n$-dimensional Gaussian posterior distribution with
mean vector $\bm{\mu}_{\mathbf{c}}$ and covariance matrix $\bm{\Sigma}_{\mathbf{c}}$ from Eq.~(\ref{eq:Baynew6}), each sample giving rise to a single set of parameters $\mathbf{c}$, and then plotting 
$\tilde{y}$ (a given realization of Eq.~(\ref{eq:predi2})); in other words, the results shown in 
the left panel have nothing to do with the posterior predictive distribution of Eq.~(\ref{eq:predi5}). 

Finally, note that some references (somewhat more carefully than the quote given above)
claim that the left panel shows
``samples from the posterior predictive distribution induced by the parameter
posterior'': this is obscure at best and misleading at worst. The 
posterior predictive distribution integrates out the parameters $\mathbf{c}$,
as you can see in Eq.~(\ref{eq:predi5}) or Eq.~(\ref{eq:predi1}); it is
therefore hard to see how 10 samples from the posterior distribution can play
the same role as a full integral over the entire posterior distribution.

{\bf Correction \#6} \textit{The natural definition of a predictive distribution,
in keeping with what is done in least-squares fitting, is to include no ad hoc
noise term. For the case of a Gaussian posterior distribution, this implies 
that the posterior predictive distribution is given by 
an expression that you are not likely to encounter in a textbook (an integral over all parameter values of the product of a one-dimensional Dirac delta function and a multivariate Gaussian). 
On a different note, 
if you wish to draw samples from the posterior predictive distribution then you have to acknowledge that you are dealing with a univariate
Gaussian and the results will therefore look jagged.
}

\section{Summary and conclusion}

In this article we encountered a sampling of mistaken claims in the introductory
literature on data analysis, machine learning, or computational science \& engineering, all
having to do with frequentist and Bayesian inference. The themes
touched upon ranged from maximum-likelihood estimation to the 
meaning of the $\chi^2$ statistic, Gaussian distributions 
(and the corresponding confidence regions) in one or many dimensions,
the question of random vs non-random in frequentist vs Bayesian regression, 
as well as the derivation of (and samples from) the posterior predictive distribution.
With the exception of Mistake \#5 all of these were very practical;
some were computational questions (e.g., is $\hat{\mathbf{c}}$ equal to 
$\mathbf{c}_{\star}$  or what does $1\sigma$ mean in many dimensions),
but nearly all of them involved a fair bit of interpretation (e.g., of the magnitude of $\chi^2$, of the meaning of a confidence interval, of what is random, or of what constitutes a prediction). 
While we tried above, where possible, to explain why a given misconception
arose in the first place, we consciously stay away from the question of why these 
issues have been propagating through the introductory (or sometimes not-so-introductory) literature. 

Before closing, we note that we are well aware that 
debunking itself has its pitfalls: there is certainly the danger that 
a careless reader may replace one set of claims
that drops from the sky with another one of the same provenance.  
Put another way, while it is reasonable to begin a myth-busting process
having Alexander Pope's ``a little learning is a dangerous thing'' in mind,
this has to 
be balanced against Heraclitus' ``the learning of many things does not teach understanding''. Our motivation in collecting these misconceptions and discussing
them in some detail is broader than the specific mistakes encountered above;
we wish to promote understanding from first principles and this is something that
one has to arrive at for oneself.  
Even so, we hope that our article is read and appreciated by data-analysis instructors
and may therefore contribute toward improved teaching and learning;
similarly, we hope that our pointing out of previously unperceived subtleties
may help elevate the level of the discussion in related research works.

\

\section*{Acknowledgments}
This work was supported in part by the Natural Sciences and Engineering Research Council (NSERC) of Canada and the Canada Foundation for Innovation (CFI). 
Computational resources were provided by SHARCNET and NERSC.


\begin{thebibliography}{99}

\bibitem{GezerlisWilliams} A. Gezerlis and M. Williams, ``Six textbook mistakes in computational physics'', Am. J. Phys. \textbf{89}, 51-60, (2021).

\bibitem{Gezerlis} A. Gezerlis, \textit{Numerical Methods in Physics with Python}, 2nd ed. (Cambridge University Press, Cambridge, 2023). 

\bibitem{Barlow} R. J. Barlow, \textit{Statistics: A Guide to the Use of Statistical Methods in the Physical Sciences} (John Wiley \& Sons, New Jersey, 1989).

\bibitem{Beers} K. J. Beers, \textit{Numerical Methods for Chemical Engineering} (Cambridge University Press, Cambridge, 2007).

\bibitem{Bertsekas} D. P. Bertsekas and J. N. Tsitsiklis, \textit{Introduction to Probability}, 2nd ed. (Athena Scientific, Massachusetts, 2008).

\bibitem{Bevington} P. R. Bevington and D. K. Robinson, \textit{Data Reduction and Error Analysis in the Physical Sciences}, 3rd ed. (McGraw-Hill, New York, 2003).

\bibitem{Bishop} C. M. Bishop, \textit{Pattern Recognition and Machine Learning} (Springer, Berlin, 2006). 

\bibitem{Bohm} G. Bohm and G. Zech, \textit{Introduction to Statistics and Data Analysis for Physicists}, 3rd ed. (Verlag Deutsches Elektronen-Synchrotron, Hamburg, 2017).

\bibitem{Boudreau} J. F. Boudreau and E. S. Swanson, \textit{Applied Computational Physics} (Oxford
University Press, Oxford, 2018).

\bibitem{Burden} R. L. Burden, D. J. Faires, and A. M. Burden, \textit{Numerical Analysis}, 10th ed. (Cengage 
Learning, Massachusetts, 2015).

\bibitem{Chapra} S. C. Chapra and R. P. Canale, \textit{Numerical Methods for Engineers}, 7th ed. (McGraw-Hill, New York, 2014).

\bibitem{Degroot} M. H. DeGroot and M. J. Schervish, \textit{Probability and Statistics}, 4th ed. (Addison-Wesley,
Massachusetts, 2012).

\bibitem{Deisenroth} M. P. Deisenroth, A. A. Faisal, and C. S. Ong, \textit{Mathematics for Machine Learning} (Cambridge University Press, Cambridge, 2020).

\bibitem{DeVries} P. L. DeVries, \textit{A First Course in Computational Physics} (John Wiley \& Sons, New Jersey, 1994).

\bibitem{Gilat} A. Gilat and V. Subramaniam, \textit{Numerical Methods for Engineers and Scientists},
3rd ed. (John Wiley \& Sons, New Jersey, 2013).

\bibitem{Gould} H. Gould, J. Tobochnik, and W. Christian, \textit{An Introduction to Computer Simulation Methods}, Rev. 3rd ed. (CreateSpace, California, 2017).

\bibitem{Hamming} R. W. Hamming, \textit{Numerical Methods for Scientists and Engineers},
2nd ed. (McGraw-Hill, 1973).

\bibitem{Jiang} H. Jiang, \textit{Machine Learning Fundamentals} (Cambridge University Press, Cambridge, 2021).

\bibitem{Kahaner} D. Kahaner, C. Moler, and S. Nash, \textit{Numerical Methods and Software} (Prentice Hall, New Jersey, 1989).

\bibitem{Kiusalaas} J. Kiusalaas, \textit{Numerical Methods for Engineers with Python 3} (Cambridge
University Press, Cambridge, 2013).

\bibitem{Koonin} S. Koonin and D. C. Meredith, \textit{Computational Physics} (Addison-Wesley, Massachusetts, 1990).

\bibitem{Landau}   R. H. Landau, M. J. P\'aez, and C. C. Bordeianu,
     \textit{Computational Physics}, 3rd ed. (Wiley-VCH, New Jersey, 2015).

\bibitem{Lyons} L. Lyons, \textit{Statistics for nuclear and particle physicists} (Cambridge University Press, Cambridge, 1986).

\bibitem{Mandel} J. Mandel, \textit{The Statistical Analysis of Experimental Data} (John Wiley \& Sons, New Jersey, 1964).

\bibitem{Mathews} J. Mathews and R. L. Walker, \textit{Mathematical Methods of Physics}, 2nd ed. (Pearson, London, 1971).

\bibitem{Murphy} K. P. Murphy, \textit{Probabilistic Machine Learning} (The MIT Press, Massachusetts, 2022).

\bibitem{Press} 
W. H. Press, S. A. Teukolsky, W. T. Vetterling, and B. P. Flannery, 
\textit{Numerical Recipes in Fortran}, 2nd ed. (Cambridge
University Press, Cambridge, 1992).

\bibitem{Pruneau} C. A. Pruneau, \textit{Data Analysis Techniques for Physical Scientists} (Cambridge
University Press, Cambridge, 2017).

\bibitem{Rice} J. A. Rice, \textit{Mathematical Statistics and Data Analysis} (Duxbury, California, 2007).

\bibitem{Roe} B. P. Roe, \textit{Probability and Statistics in the Physical Sciences}, 3rd ed. (Springer, Berlin, 2020).

\bibitem{Rogers} S. Rogers and M. Girolami, \textit{A First Course in Machine Learning}, 2nd ed. (CRC Press, Florida, 2017).

\bibitem{Sirca} S. {\v{S}}irca and M. Horvat, \textit{Computational Methods in Physics}, 2nd ed. (Springer, Berlin, 2018).

\bibitem{Sivia} D. S. Sivia and J. Skilling, \textit{Data Analysis: A Bayesian Tutorial}, 2nd ed. (Oxford University Press, Oxford, 2006).

\bibitem{Theodoridis} S. Theodoridis, \textit{Machine Learning: A Bayesian and Optimization Perspective}, 2nd ed. (Academic Press, London 2020).

\bibitem{Thompson} W. J. Thompson, \textit{Computing for Scientists and Engineers}
(John Wiley \& Sons, New Jersey, 1992).

\bibitem{Wong} S. S. M. Wong, \textit{Computational Methods in Physics and Engineering}, 2nd ed. (World Scientific, Singapore, 1997).

\bibitem{Zielesny} A. Zielesny, \textit{From Curve Fitting to Machine Learning}, 2nd ed. (Springer, Berlin, 2016).



\bibitem{Casella} G. Casella and R. L. Berger, \textit{Statistical Inference}, 2nd ed. (Duxbury, California, 2002).

\bibitem{Wasserman} L. Wasserman, \textit{All of Statistics: A Concise Course in Statistical Inference} (Springer, Berlin, 2004), chapter~9.

\bibitem{Kendall} M. G. Kendall and A. Stuart, \textit{The Advanced Theory of Statistics, Volume 2} (Hafner Publishing Company, New York, 1961), p. 40.

\bibitem{Vehtari} A. Vehtari, D. P. Simpson, Y. Yao, and A. Gelman, ``Limitations of `Limitations of Bayesian leave-one-out cross-validation for
model selection' '', Comput. Brain Behav. \textbf{2}, 22-27, (2019).

\bibitem{Gregory} P. C. Gregory, \textit{Bayesian Logical Data Analysis for the Physical Sciences} (Cambridge
University Press, Cambridge, 2005), p. 280.

\bibitem{Feynman} D. L. Goodstein, ``Richard P. Feynman, Teacher'', Physics Today, \textbf{42}, 70-75 (February 1989).

\bibitem{Andrae} R. Andrae, T. Schulze-Hartung, and P. Melchior, ``Dos and don’ts of reduced
chi-squared'',  arXiv:1012.3754.

\bibitem{Robert} C. P. Robert, \textit{The Bayesian Choice}, 2nd ed. (Springer, Berlin, 2007), p. 350.

\bibitem{Morey} R. D. Morey, R. Hoekstra, J. N. Rouder, M. D. Lee,
and E.-J. Wagenmakers, ``The fallacy of placing confidence in confidence intervals'', 
Psychon. Bull. Rev. \textbf{23}, 103-123, (2016).

\bibitem{Bajorski} P. Bajorski, \textit{Statistics for Imaging, Optics, and Photonics} 
(John Wiley \& Sons, New Jersey, 2012), p. 166.

\bibitem{Diaconis} P. Diaconis and B. Skyrms, \textit{Ten Great Ideas About Chance} (Princeton University Press, New Jersey, 2018).

\bibitem{Gelman} A. Gelman, J. B. Carlin, H. S. Stern, D. B. Dunson, A. Vehtari, and D. B. Rubin, \textit{Bayesian Data Analysis}, 3rd ed. (CRC Press, Florida, 2013).

\bibitem{Fuchs} C. A. Fuchs and R. Schack, ``QBism and the Greeks: why a quantum state does not represent an element of physical reality'', Phys. Scr. \textbf{90}, 015104 (2015).

\bibitem{BernardoSmith} J. M. Bernardo and A. F. M. Smith, \textit{Bayesian Theory} (John Wiley \& Sons, New Jersey, 2000), p. 236.

\bibitem{Sprenger} J. Sprenger and S. Hartmann, \textit{Bayesian Philosophy of Science} 
(Oxford
University Press, Oxford, 2019), p. 28.

\bibitem{Bernardo} J. M. Bernardo, ``Interpretation of Electoral Results:
A Bayesian Analysis'', Proceedings of Teias Matem\'aticas, 63-75 (2004).

\bibitem{Jaynes} E. T. Jaynes, \textit{Probability Theory: The Logic of Science} 
(Cambridge
University Press, Cambridge, 2003), p. 108.

\bibitem{Williams} B. Williams, \textit{Moral Luck: Philosophical Papers 1973-1980} (Cambridge
University Press, Cambridge, 1982), p. 18.

\end{thebibliography}
\end{document}